\documentclass[fleqn,usenatbib]{mnras}

\usepackage{newtxtext,newtxmath}

\usepackage[T1]{fontenc}
\usepackage[utf8]{inputenc}
\usepackage[english]{babel}
\usepackage{lscape}
\DeclareRobustCommand{\VAN}[3]{#2}
\let\VANthebibliography\thebibliography
\def\thebibliography{\DeclareRobustCommand{\VAN}[3]{##3}\VANthebibliography}

\usepackage{graphicx}	
\usepackage{amsmath}	
\usepackage{dblfloatfix}
\usepackage{caption}
\usepackage{subcaption}
\usepackage{ulem}

\newcommand{\angstrom}{\text{\normalfont\AA}}

\title[Smaller quasar black hole masses]{Systematically smaller single-epoch quasar black hole masses using a radius-luminosity relationship corrected for spectral bias}

\author[J. Maithil et al.]{Jaya Maithil,$^{1}$\thanks{E-mail: jmaithil@uwyo.edu}
Michael S. Brotherton,$^{1}$ Ohad Shemmer,$^{2}$ Pu Du,$^{3}$ Jian-Min Wang,$^{3,4,5}$
\newauthor{ Adam D. Myers,$^{1}$ Jacob N. McLane,$^{1}$ Cooper Dix,$^{2}$ and Brandon M. Matthews$^{2}$}
\\
$^{1}$Department of Physics and Astronomy, University of Wyoming, Laramie, WY 82071, USA \\
$^{2}$Department of Physics, University of North Texas, Denton, TX 76203, USA \\
$^{3}$Key Laboratory for Particle Astrophysics, Institute of High Energy Physics, Chinese Academy of Sciences, 19B Yuquan Road, Beijing 100049,\\ People’s Republic of China \\
$^{4}$National Astronomical Observatories of China, Chinese Academy of Sciences, 20A Datun Road, Beijing 100020, People’s Republic of China \\
$^{5}$School of Astronomy and Space Science, University of Chinese Academy of Sciences, 19A Yuquan Road, Beijing 100049, People’s Republic of China \\}

\date{Accepted 2022 June 17. Received 2022 June 16; in original form 2021 May 1}

\pubyear{2022}

\begin{document}
\label{firstpage}
\pagerange{\pageref{firstpage}--\pageref{lastpage}}
\maketitle

\begin{abstract}

Determining black hole masses and accretion rates with better accuracy and precision is crucial for understanding quasars as a population. These are fundamental physical properties that underpin models of active galactic nuclei. A primary technique to measure the black hole mass employs the reverberation mapping of low-redshift quasars, which is then extended via the radius-luminosity relationship for the broad-line region to estimate masses based on single-epoch spectra. An updated radius-luminosity relationship incorporates the flux ratio of optical Fe {\sc ii} to H$\beta$ ($\equiv \mathcal{R}_{\rm Fe}$) to correct for a bias in which more highly accreting systems have smaller line-emitting regions than previously realized. In this current work, we demonstrate and quantify the effect of using this Fe-corrected radius-luminosity relationship on mass estimation by employing archival data sets possessing rest-frame optical spectra over a wide range of redshifts. We find that failure to use a Fe-corrected radius predictor results in overestimated single-epoch black hole masses for the most highly accreting quasars. Their accretion rate measures ($L_{\rm Bol}/ L_{\rm Edd}$ and $\dot{\mathscr{M}}$), are similarly underestimated. The strongest Fe-emitting quasars belong to two classes: high-z quasars with rest-frame optical spectra, which given their extremely high luminosities, require high accretion rates, and their low-z analogs, which given their low black holes masses, must have high accretion rates to meet survey flux limits. These classes have mass corrections downward of about a factor of two, on average. These results strengthen the association of the dominant Eigenvector 1 parameter $\mathcal{R}_{\rm Fe}$ with the accretion process.

\end{abstract}




\section{Introduction}
Black hole mass and accretion rate are arguably the two most important properties of quasars. They are vital in understanding the growth of supermassive black holes and how feedback from quasars regulates the growth of massive galaxies. In the past decade, more than 50 quasars have been discovered at $z>6$ that are claimed to have billion solar mass black holes \citep[e.g.,][]{Mortlock_etal2011,Wang_2021}. Such findings challenge our understanding of black hole formation and growth. To form a billion solar mass black hole when the universe was $\leq$ 1 Gyr old requires a massive seed black hole, $M_{\rm seed} \geq 10^3 M_{\sun}$, in an Eddington-limited accretion rate scenario \citep[e.g.,][]{Lodato&Natarajan2006, Johnson_etal2012}. Alternatively, accretion with reduced radiative efficiency, $\epsilon = 0.01-0.001$, could explain the growth in such a short cosmic time \citep[e.g.,][]{Volonteri_etal2015,Trakhtenbrot_etal2017,Davies_etal2019}. Another possible explanation for such apparently large quasar black hole masses at such 
early times could also simply be a systematic overestimation.  

For nearby (distance $<$ 300 Mpc, $z<0.06$) active galactic nuclei (AGNs), direct measurement of black hole mass is made using stellar and gas dynamics \citep[][]{Kormendy_Ho2013}. These methods are not possible for distant objects because of limited angular resolution and the fact that quasars outshine their host galaxy. Observations of the correlated variation of the continuum and photoionized broad emission lines (especially Balmer lines) in type-1 AGNs led to the development of the reverberation mapping (RM) technique to determine black hole masses \citep[e.g.,][]{Peterson1993}. Time delays between the continuum and emission-line variability, derived from multiple epochs of spectroscopy, can be used to measure the size of the broad-line region (BLR) \citep[][]{Blandford1982, Netzer&Peterson1997, Peterson1993, Peterson2014}. Combining this measurement with the emission-line velocity dispersion provides an estimate of the virialized black hole mass (see equation 1) \citep[][]{Peterson_etal2004}. Over the past two decades, RM has provided black hole mass measurements for over a hundred AGNs \citep[e.g.,][]{Bentz_Katz2015, Yu_etal2020}. However, it is impractical to apply the RM method for every AGN as it is resource-intensive. Fortunately, RM studies show a relationship between the BLR size and the continuum luminosity to be $\rm R \propto L^{\sim0.5}$ \citep{Kaspi2000, Kaspi_etal2005,Kaspi+2007, Kaspi+2021, Bentz+2006, Bentz_etal2009, Bentz2013}. RM studies based on the H$\beta$ line provide the most reliable correlations \citep[][]{Bentz2013}. Using the luminosity at 5100$\angstrom$ from single-epoch (SE) spectra, one can then predict the size of the BLR and estimate the black hole mass \citep[e.g.,][]{Laor1998, Wandel_etal1999, VP2006}.

For the highest redshift quasars, most of the optical-UV lines fall in the near-infrared part of the spectrum, for which there are fewer spectral observations. The masses of the black holes that power high-redshift quasars are therefore estimated using UV continuum luminosities and emission lines like C {\sc iv} and Mg {\sc ii}, calibrated against $\rm H\beta$ RM measurements, a procedure that introduces additional uncertainties. The offset between Mg {\sc ii} or C {\sc iv}-based black hole masses with H$\beta$-based masses is in part related to Eigenvector 1  (EV1) spectral trends \citep[e.g.,][]{Shen_etl2008, Runnoe)etal2013, Brotherton_etal2015} that appear to be correlated with Eddington ratio \citep[][]{BG1992, Boroson2002, Yuan+2003, Marziani+2001, Shen_Ho2014, Sun&Shen2015}. \cite{Shen_etl2008} demonstrated that the offset between Mg {\sc ii} and C {\sc iv}-based SE black hole mass correlates with another EV1 parameter, C {\sc iv} blueshift. \cite{Runnoe)etal2013} identified a similar bias in C {\sc iv}-based masses using the ratio of C {\sc iv} to the $\lambda$1400 feature (a blend of Si {\sc iv} + O {\sc iv}]). \cite{Brotherton_etal2015} demonstrated that an EV1 bias in reverberation-mapped AGN samples that leads to a 50$\%$ overestimation of C {\sc iv}-based masses in average quasars. Recent results from \cite{Bonta_etal2020} show that the difference between C {\sc iv}-based RM masses and SE masses anti-correlates with Eddington ratio. 

The H$\beta$ RM sample originally used to establish the R-L relationship primarily included objects with strong narrow [O {\sc iii}] emission lines. This is because [O {\sc iii}] lines are convenient for relative flux calibration \citep{Groningen+Wanders1992} and such objects also tend to have strong broad H$\beta$ line variability. The equivalent width (EW) of [O {\sc iii}] is anti-correlated with the Eddington ratio ($L_{\rm bol}/L_{\rm Edd}$, described in Section 2) \citep[][]{BG1992, Marziani+2001, Boroson2002, Shen_Ho2014}, hence, RM samples were biased toward low-accretion-rate broad-lined AGNs (i.e., Eddington ratio of a few to a few tens of percent). Recent H$\beta$ RM campaigns, such as the Super-Eddington Accreting Massive Black Hole \citep[SEAMBH;][]{Du_etal2014, Du_etal2016, Du_etal2018} and the Sloan Digital Sky Survey Reverberation Mapping projects \citep[SDSS-RM;][]{Shen_etal2015} find deviations from the canonical R-L relationship. The observed time lags are sometimes significantly smaller than predicted \citep{Du_etal2015, Du_etal2016, Grier_etal2017, Du_etal2018, Du_2019}. The offsets in the H$\beta$ SDSS-RM sample are not due to observational bias, but rather they reflect the wide variety of broad-line radii occupied by  AGNs \citep{Alvarez_etal2020}. Moreover, the offset between the observed and predicted BLR radius shows an anti-correlation with the EV1 accretion rate parameters \citep{Du_etal2018, Du_2019, Bonta_etal2020}. Even the Mg {\sc ii} and C {\sc iv} RM  samples demonstrate similar offsets in BLR radius that are correlated with accretion rate parameters, suggesting that current R-L relationships should include some additional correction terms \citep{Aldama_etal2020, Bonta_etal2020}. 

The SEAMBH H$\beta$ RM sample comprises a population of highly accreting AGNs with a BLR radius up to 3-8 times smaller than predicted from the canonical R-L relationship, which implies an overestimation of SE black hole masses by the same factor. The SEAMBH RM results also establish a strong anti-correlation between the deviation from the canonical R-L relationship and the relative strength of Fe {\sc ii}, an EV1 parameter that correlates with the Eddington ratio. Using a sample of 75 RM AGNs, \cite{Du_2019} updated and tightened the R-L relationship by introducing the relative strength of Fe {\sc ii} as a predictive parameter. \cite{Yu_etal2020} provide a similar accretion-rate-based correction to the R-L relation using the strength of Fe {\sc ii}. Such a correction should be extremely significant for luminous high-redshift quasars, which are likely to be accreting at a high rate. Using the Eddington ratio distribution for a uniformly selected sample of type 1 quasars from SDSS DR7, \cite{Kelly_Shen2013} employed a flexible Bayesian technique to demonstrate that the fraction of quasars with a higher Eddington ratio becomes larger at high redshift. Therefore, the canonical R-L relationship most likely overestimates the masses of the black holes hosted by high-redshift quasars. Similarly, the dimensionless accretion rate parameter ($\dot{\mathscr{M}}$) and Eddington ratio, which are inversely proportional to black hole mass, are likely to be underestimated. Even this underestimated $\dot{\mathscr{M}}$ results in a large population of super-Eddington quasars, $\dot{\mathscr{M}}>3$, which necessitates the use of an accretion rate corrected R-L relationship (see Figure \ref{fig:Figure4} in Section 4). 

This paper adopts the \cite{Du_2019} R-L relationship to quantify this effect using archival data of low and high-redshift quasars. Our results demonstrate that for objects with large $\mathcal{R}_{\rm Fe}$, the SE method adopting the canonical R-L relation significantly overestimates their $\rm H\beta$-based masses and underestimates their accretion rates by factors of a couple to several. In section 2, we explain the method to determine black hole mass and accretion rate parameters with the new and canonical R-L relationship. Section 3 describes the low and high redshift samples used and summarizes the quantities used to estimate black hole mass. We discuss our results regarding the black hole mass and accretion rate in section 4.1 and the correlation between $\mathcal{R}_{\rm Fe}$ and accretion rate parameters in section 4.2, followed by additional discussion and conclusions in sections 5 and 6, respectively. Throughout the paper, we adopt a cosmology with $\rm H_0 = 70~km~s^{-1}~Mpc^{-1}$, $\rm \Omega_\Lambda = 0.7$ and $\Omega_m = 0.3$.

\section{Black hole mass and accretion rate}
Black hole masses ($M_{\rm BH}$) are estimated using the following relationship
\begin{equation}
   M_{\rm BH} = \textit{f}~ \left(\frac{R_{\rm BLR}\Delta V^2}{G}\right).
    \label{eq:eq1}
\end{equation}
This equation assumes the virialized motion of BLR clouds under the gravitational potential of the central black hole \citep[e.g.,][]{Wandel_etal1999, Peterson_etal2004}. The expression in parenthesis is called virial product. The full-width at half maximum (FWHM) or line dispersion ($\sigma_{\rm line}$) of broad emission lines like $\rm H\beta$ sets the velocity ($\Delta V$) assuming Doppler broadening. The size of the BLR, $R_{\rm BLR}$, can then be determined by multiplying the time lag ($\Delta \tau$) between emission-line and continuum variability, determined by reverberation mapping, by the speed of light ($R_{\rm BLR} = c \Delta \tau $). The virial coefficient, $f$, accounts for the unknown geometry, kinematics, and inclination of the BLR. Although its value differs from one AGN to another, a mean value of $f$ is obtained empirically by calibrating RM mass against mass predicted by the $M-\sigma_{\star}$ relation seen in quiescent galaxies \citep[][and many others since]{Onken+2004}. Here  we focus on H$\beta$-based  SE virial black hole mass and adopt FWHM of H$\beta$ as the measure of $\Delta V$.

The monochromatic luminosity at 5100$\angstrom$ in $\rm erg~ s^{-1}$, $\lambda \it{L}_{\lambda, 5100\angstrom}$ (hereafter $L_{5100 \angstrom}$), serves as a proxy for $R_{\rm BLR}$. We use two R-L relationships: (1) the canonical R-L relationship established by \cite{Bentz2013}

\begin{equation}
   \rm log\ (\it{R}_{\rm H\it{\beta}}/ {\rm lt - days}) = K + \mu~{\rm log \left(\it{L}_{5100\angstrom} /\rm 10^{44}~ erg~ s^{-1} \right)},
   \label{eq:eq2}
\end{equation}
where, $K = 1.527\pm0.031$ \& $\mu = 0.533\pm 0.035$, and (2) an accretion rate corrected R-L relationship established by \citet{Du_2019}
\begin{equation}
   \rm log\ (\it{R}_{\rm H\it{\beta}}/ \rm lt - days) = \alpha + \beta log\ \ell_{44} + \gamma \mathcal{R}_{Fe},\vspace{-2ex}
   \label{eq:eq3}
\end{equation}
where, $\ell_{44} = L_{5100\angstrom}$/$\rm 10^{44}\ erg\ s^{-1}$, $\rm \alpha =  1.65\pm0.06; \beta = 0.45\pm0.03; \gamma = -0.35\pm0.08$. It takes into account the relative strength of Fe {\sc ii}, $\mathcal{R}_{\rm Fe}$, that is known to correlate with EV1. $\mathcal{R}_{\rm Fe}$ is defined as the ratio of flux ($F$) or rest-frame equivalent width (EW) between Fe {\sc ii} and $\rm H\beta$, i.e., $\mathcal{R}_{\rm Fe} = F$(Fe {\sc ii})/$F(\rm H\beta) \approx$ EW(Fe {\sc ii})/EW$(\rm H\beta)$. 
A higher $\mathcal{R}_{\rm Fe}$ value leads to a systematically smaller $R_{\rm BLR}$ estimate and is likely associated with a higher accretion rate \citep[][and our discussion later in this paper]{Du_2019}.

We estimate the SE virial black hole mass by determining $R_{\rm BLR}$ from the R-L relationships, using FWHM of H$\beta$ as the velocity term, and a virial coefficient $f=1.5$ in equation \ref{eq:eq1}. Our choice of the virial coefficient is consistent with the empirical mean value of $f$ obtained by calibrating FWHM-based RM black hole mass from rms spectra with the $M-\sigma_{\star}$ relation. We adopted the \citet{Ho&Kim2014} value of $f = 1.5 \pm 0.4$ for AGNs in classical bulges and ellipticals that have black hole mass greater than $10^7 M_{\sun}$. Recent work by \cite{Yu+2020, Yu+2019} also finds $f ~=1.51\pm0.20$ for low-redshift RM AGNs in classical bulges and ellipticals. Henceforth, we shall refer to the SE black hole mass estimated using equation \ref{eq:eq2} as the ``canonical'' black hole mass, $M_{\rm H\beta, canonical}$ and the one using equation \ref{eq:eq3} as the ``Fe-corrected'' black hole mass, $M_{\rm H\beta, Fe-corrected}$. Note that we adopt $f=1.5$ for both canonical and Fe-corrected black hole mass estimates. Therefore, the $\mathcal{R}_{\rm Fe}$ factor in the \citet{Du_2019} R-L relationship (equation \ref{eq:eq3}) dominates the difference between $M_{\rm H\beta, canonical}$ and $M_{\rm H\beta, Fe-corrected}$.

We calculate two relative accretion rate parameters based on luminosity and black hole mass. First, there is the dimensionless accretion rate parameter, $\dot{\mathscr{M}}$, derived from the \cite{ShakuraandSunyaev1973} thin accretion disk model for which
\begin{equation}
    \dot{\mathscr{M}} = 20.1 \left( \ell_{44}/\cos i \right)^{3/2} m_7^{-2},\vspace{-1ex}
    \label{eq:eq4}
\end{equation}
where, $m_7 = M_{\rm BH}$/$10^7 M_{\sun}$, and $i$ is inclination angle to the line of sight \citep{Wang2014, Du_etal2014, Du_etal2018, Du_2019}. We take $\cos i = 0.75$, an average for type 1 AGNs, for our calculation. Second, we have the Eddington ratio, ($L_{\rm bol}/L_{\rm Edd}$) expressed as the ratio of the bolometric luminosity ($L_{\rm Bol}$) to the Eddington luminosity ($L_{\rm Edd} = 1.5 \times 10^{45}~ m_7$ erg s$^{-1}$). We calculate $L_{\rm Bol}$ using the \cite{Richards_etal2006} relation, $L_{\rm Bol}=9.26~L_{5100 \angstrom}$.
\begin{equation}
    L_{\rm bol}/L_{\rm Edd} = \frac{9.26~L_{5100 \angstrom}}{1.5 \times 10^{45}~ m_7}
    \label{eq:eq4b}
\end{equation}
The bolometric luminosity may saturate in the case of quasars with super-Eddington accretion owing to the photon trapping effect in their slim accretion disks \citep{Wang+1999, Mineshige2000}. This, in principle, may make $\dot{\mathscr{M}}$ a better predictor of accretion rate, although the two are highly correlated for quasars, as we will show.

For the flux-to-luminosity conversion we use $\lambda L_{\lambda}[\rm erg~s^{-1}]$ = 4 $\pi D_{\rm L}^2 f_{\lambda} (1+z) \lambda$, where $f_{\lambda}$ is the monochromatic flux at the rest-wavelength ($\lambda$) in units of $\rm erg~s^{-1}~cm^{-2}~\angstrom^{-1}$, $D_{\rm L}$ is the luminosity distance, and $z$ is the redshift \citep[][]{Hogg2002}.

\section{Archival data sets and measurements}
To characterize the effect of using the new R-L relationship (equation 3) on H$\beta$-based black hole mass estimates, and hence also estimates of accretion rate, we selected archival samples and associated catalogs that provide: a) flux or EW of Fe {\sc ii} between rest-frame 4435-4685 $\angstrom$ and the broad H$\beta$ component to calculate $\rm \mathcal{R}_{Fe}$, b) flux or luminosity at rest-frame 5100 $\angstrom$ to estimate the BLR size from the new R-L relationship, and c) FWHM H$\beta$ to provide a proxy for velocity dispersion. 

For low-redshift quasars ($z < 0.7$), we used only the catalog of \cite{Shen_2011}, which is highly complete with uniform measurements of the quantities we need for objects in the Sloan Digital Sky Survey Data Release 7 \citep[SDSS DR7]{SDSSDR7_2009}.  The \citet{Shen_2011} catalog contains a total of 105,783 quasars. We applied a conservative S/N and redshift  cutoff to obtain a sample unbiased by poor-quality spectra in the H$\beta$ region. Our selection criteria include a median $\rm S/N$ per pixel in the $\rm H\beta$ region greater than 20, redshift $z<0.7$, and non-zero measurements of EW and FWHM $\rm H\beta$. Our choice of $\rm S/N>20$ per pixel eliminates unreliable line width measurements and reduces the  formal uncertainties in SE masses to a minimum \citep{Denney+2009}. Given the automatic nature of spectral fits in \cite{Shen_2011}, some individual measurements are bad. So we applied a $\rm S/N>3$ cut on the H$\beta$ line measurements as an additional quality control check. We eliminated three targets, SDSS J094927.67+314110.0, SDSS J105528.80+312411.3, and SDSS J151036.74+510854.6 because of erroneous measurements due to incorrect redshifts\footnote{While inspecting the SDSS spectrum of outliers in Figure \ref{fig:Figure6a}, \ref{fig:Figure6} \& \ref{fig:Figure7}, these three quasars had wrong redshifts assigned to them.}, giving a total of 3309 quasars. 

\cite{Shen_2011} did not correct the cataloged 5100 $\angstrom$ luminosities for host-galaxy contamination. We applied a correction when log$[L_{5100\angstrom} \rm /(erg~ s^{-1})]<45.053$, using equation 1 of \cite{Shen_2011}. Recently, \cite{Bonta_etal2020} defined a host-galaxy light correction based on the luminosity of the H$\beta$ line, L(H$\beta$). We tested the effect of our choice of using \cite{Shen_2011} host-galaxy correction method against the method described by \cite{Bonta_etal2020} for the canonical and Fe-corrected black hole mass estimates. There is essentially no change in the mass difference distribution between the two different methods when correcting $L_{5100\angstrom}$ for the low-luminosity sub-sample (log $[L_{5100\angstrom} \rm /(erg~ s^{-1})] <$ 45.0). The mean mass differences are 0.14 dex with the \citeauthor{Shen_2011} correction, 0.13 dex with the \citeauthor{Bonta_etal2020} correction, and the standard deviations are 0.19 dex and 0.18 dex, respectively. There are systematic differences in the host-galaxy correction method defined by \citeauthor{Bonta_etal2020} and \citeauthor{Shen_2011}. On average, \citeauthor{Bonta_etal2020} method underestimates $L_{5100\angstrom}$ by a factor of 1.28 compared to \citeauthor{Shen_2011} method. For highly luminous quasars (log [$L_{5100\angstrom} /\rm (erg ~s^{-1})]$ > 45.0), the mean underestimation in $L_{5100\angstrom}$ by \citeauthor{Bonta_etal2020} method is a factor of 1.32 compared to \citeauthor{Shen_2011}, with 40\% of quasars underestimated by a factor of 1.32-13.18. Note that the Shen et al. method is not sensitive to any host galaxy emission when the spectral absorption features disappear; hence this method underestimates the host-galaxy correction for the most luminous quasars.
Even though the \citeauthor{Bonta_etal2020} $L(\rm H\beta$)-$L_{5100\angstrom}$ correlation is relatively tight, it is based on a small sample, and there may be issues extrapolating to higher $L_{5100\angstrom}$. There is an additional concern that EW H$\beta$ correlates with $\mathcal{R}_{\rm Fe}$ (P < 1\%; \cite{BG1992}) and using an $L_{5100\angstrom}$ based on $L(\rm H\beta$) might bias Fe-corrected black hole mass estimates.

For high-redshift quasars, the H$\beta$ region falls in the near-infrared (IR). A good signal-to-noise (S/N) near-IR spectrum of a low-luminosity high-redshift quasar requires an exorbitantly large amount of observing time on most telescopes. As a result, archival samples at high redshift predominantly contain high-luminosity quasars. Only a handful of samples tabulate the Fe {\sc ii} measurement essential for our study. Our analysis includes all the high-redshift samples we found that provide good spectral measurements required to calculate canonical and Fe-corrected black hole mass. These high-redshift samples are a near-IR follow-up of quasars with previous rest-frame UV spectral observation selecting targets based primarily on two criteria: 1) high S/N ratio in the spectral region containing UV emission lines like C {\sc iv} and N {\sc v} and 2) redshifts for which H$\beta$ falls in unobscured near-IR spectral bands (i.e., JHK bands). Appendix \ref{app:app1} provides more detailed information about the individual sample selection of the high-redshift samples from the work of
\cite{Shen2016}, the two-part series by \cite{Shemmer2004} \& \cite{Netzer2004} (hereafter SN2004), and \cite{Sulentic2017}.

\begin{figure}
   \includegraphics[trim=0.1cm 0.3cm 0.7cm 0.7cm,clip,width=0.8\columnwidth]{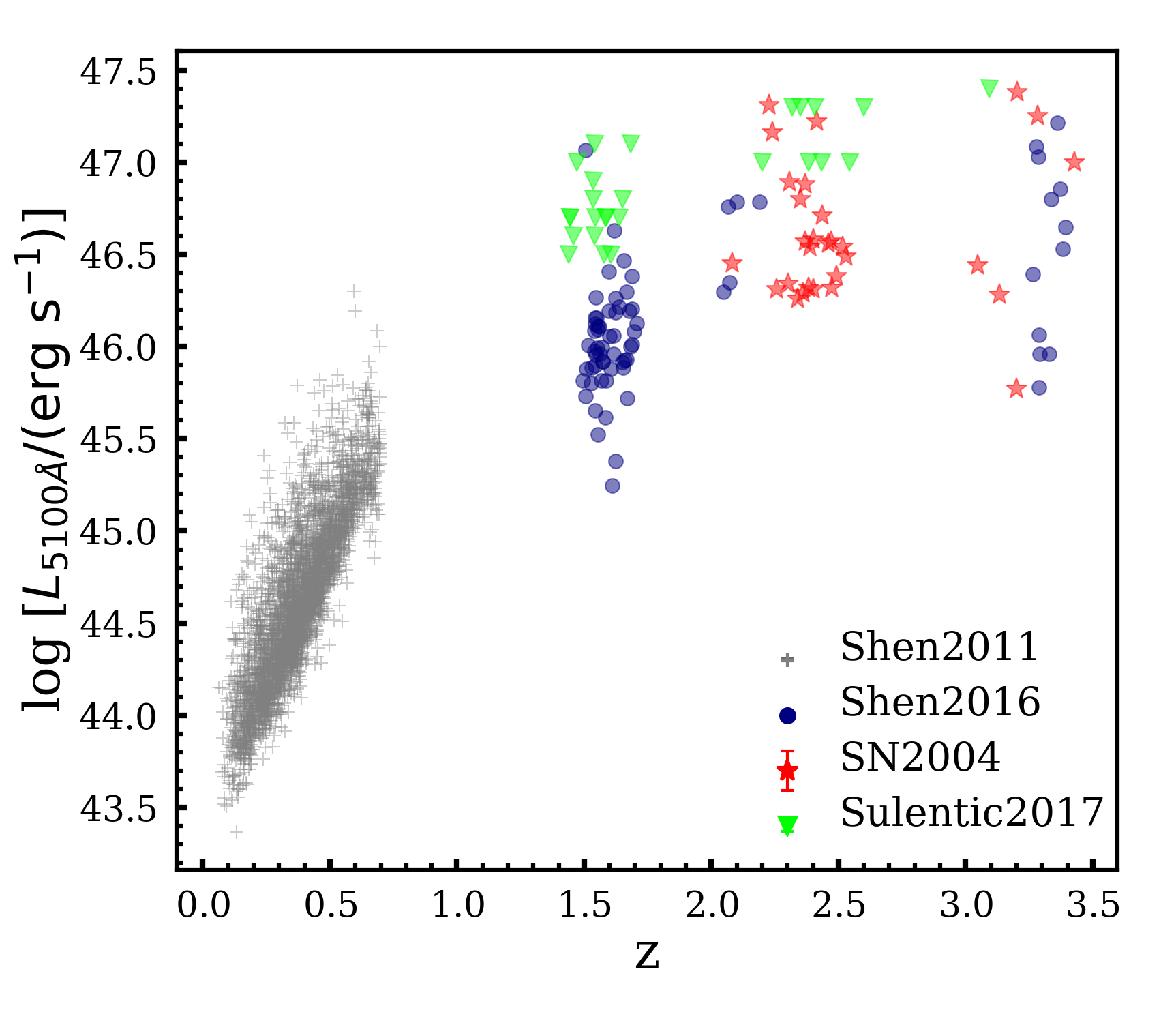}
   \caption{Redshift vs log $L_{5100\angstrom}$. The plot shows the redshift-luminosity space spanned by our quasar samples.}
   \label{fig:Figure1a}
\end{figure}

Table \ref{tab:sample} lists the name of the samples we use, their total number of objects, and their redshift range.  \cite{Shen2016} contains 74 quasars in the redshift range $1.5\leq z\leq3.5$. We eliminated one quasar, J0810+0936, with a strangely large uncertainty reported for its luminosity measurement (log [$L_{5100\angstrom}/ \rm (erg~ s^{-1})]= 46.30\pm23.16$). \cite{Shen2016} report rest-frame EWs, which we used to calculate $\mathcal{R}_{\rm Fe}$, and include 71 quasars with $\mathcal{R}_{\rm Fe}<3$. SN2004 consists of 29 quasars in the redshift range $2\leq z \leq3.5$. We used the tabulated systemic redshift, 5100$\angstrom$ luminosity and best-fit FWHM H$\beta$ values from \cite{Shemmer2004}, and  $\rm \mathcal{R}_{Fe}$ measurements given by \cite{Netzer2004}. \cite{Sulentic2017} catalog the properties of a sample of 28 quasars with $1.4\leq z\leq3.1$, including two weak-line quasars (HE0359-3959, HE2352-4010) that we eliminate for our analysis\footnote{We include the two weak-line quasars from \cite{Sulentic2017} and two others from \cite{Shemmer2010} in Appendix \ref{app:app3}}. We used their tabulated $\rm \mathcal{R}_{Fe}$ measurements and FWHM H$\beta$ of the broad component obtained from the spectral fit analysis.
Our final combined sample provides redshift coverage $0.05< z< 3.6$ and represents a typical range of luminosity seen in low and high-redshift quasars (Figure \ref{fig:Figure1a}).

\begin{table}
 \caption{Data sets used in our analysis}
 \label{tab:sample}
 \begin{tabular}{llll}
 \hline
Ref & Sample & Sample size & Redshift\\
 \hline
 \citet{Shen_2011} & Shen2011	&	3309	&	$< 0.7$\\
 \citet{Shen2016} & Shen2016	&	71	&	1.5-3.5\\
 \citet{Shemmer2004}, & SN2004	&	29	&	2-3.5\\
 \citet{Netzer2004} & & & \\
 \citet{Sulentic2017} & Sulentic2017	&	26	&	1.4-3.1\\
 \hline
 \end{tabular}
\end{table}
We calculated $M_{\rm H\beta, canonical}$ for all samples by using the canonical R-L relation (equation \ref{eq:eq2}), with FWHM H$\beta$ as a proxy for the velocity, and $f = 1.5$ in equation \ref{eq:eq1}. All of these literature sources except \cite{Sulentic2017} provide EWs, and use the EW ratio as $\mathcal{R}_{\rm Fe}$. \cite{Netzer2004} mention that the flux ratio is the same as the EW ratio used to define $\mathcal{R}_{\rm Fe}$ by \cite{BG1992}. \cite{Sulentic2017} use line flux ratios to calculate $\mathcal{R}_{\rm Fe}$, and note that this is equivalent to using EW ratios. We used the \cite{Du_2019} R-L relation (equation \ref{eq:eq3}), with FWHM H$\beta$ as a proxy for the velocity and $f = 1.5$ in equation \ref{eq:eq1}, to calculate $M_{\rm H\beta, Fe-corrected}$ for each sample. For \citet{Shen_2011} quasars, we used the luminosity corrected for host-galaxy contamination to estimate both $M_{\rm H\beta, canonical}$ and $M_{\rm H\beta, Fe-corrected}$.

For \cite{Shen_2011} and \cite{Shen2016} samples, we propagated the measurement uncertainties in FWHM H$\beta$, $L_{5100\angstrom}$, EW H$\beta$ and EW Fe {\sc ii} to calculate the error in the derived quantities like $\mathcal{R}_{\rm Fe}$, $R_{\rm BLR}$, $M_{\rm BH}$, $L_{\rm Bol}/L_{\rm Edd}$, and $ \dot{\mathscr{M}}$. For other samples, we made a few assumptions to estimate measurement uncertainties. SN2004 provide the uncertainty in FWHM H$\beta$ as the difference between direct and best-fit measures for each source and quotes an average uncertainty of 25\% on $L_{5100\angstrom}$. We assumed an uncertainty of 24\% for $\mathcal{R}_{\rm Fe}$ based on the mean uncertainty reported by \cite{McIntosh+1999} with similar data quality and spectral resolution. \cite{Sulentic2017} list uncertainty in $5100\angstrom$ flux but do not provide uncertainties in spectral measurements. As \cite{Sulentic2017} obtained the spectra from the parent samples of \cite{Sulentic+2004, Sulentic+2006} and \cite{Marziani+2009} and redid the analysis; their spectral measurements are consistent but not identical. So we obtained the relative error in FWHM H$\beta$, EW H$\beta$, and EW Fe {\sc ii} for each target in \cite{Sulentic2017} from the parent samples to estimate uncertainties in derived quantities. Our error propagation also includes uncertainties in the virial coefficient ($f$) and the coefficients in the two R-L relationships. We list the mean measurement uncertainties in the derived quantities for each sample in Table \ref{tab:error} and show them as typical error bars in our figures.

\begin{table*}
 \caption{Mean measurement uncertainties on the derived quantities for each sample}
 \label{tab:error}
 \begin{tabular}{l|ccccc}
 \hline
  Quantities & & Mean measurement error & \\
  \hline
	&	Shen2011	&	Shen2016	&	SN2004	&	Sulentic2017	\\
  \hline
$R_{\rm H\beta, canonical}$	&	0.04 dex	&	0.08 dex	&	0.11 dex	&	0.11 dex	\\
$M_{\rm H\beta, canonical}$	&	0.15 dex	&	0.20 dex	&	0.27 dex	&	0.17 dex	\\
$L_{\rm Bol}/L_{\rm Edd, canonical}$	&	0.15 dex	&	0.20 dex	&	0.27 dex	&	0.17 dex	\\
$\dot{\mathscr{M}}_{\rm canonical}$	&	0.30 dex	&	0.41 dex	&	0.53 dex	&	0.35 dex	\\
$\mathcal{R}_{\rm Fe}$	&	0.10 dex	&	0.15 dex	&	0.10 dex	&	0.11 dex	\\
$R_{\rm H\beta, Fe-corrected}$	&	0.10 dex	&	0.15 dex	&	0.14 dex	&	0.13 dex	\\
$M_{\rm H\beta, Fe-corrected}$	&	0.17 dex	&	0.24 dex	&	0.29 dex	&	0.19 dex	\\
$L_{\rm Bol}/L_{\rm Edd, Fe-corrected}$	&	0.18 dex	&	0.24 dex	&	0.29 dex	&	0.19 dex	\\
$\dot{\mathscr{M}}_{\rm Fe-corrected}$	&	0.35 dex	&	 0.48 dex	&	0.57 dex	&	0.37 dex	\\
 \hline
 \end{tabular}
\end{table*}

\begin{figure}
   \includegraphics[trim=0.cm 0.cm 0.5cm 2.cm,clip,width=0.9\columnwidth]{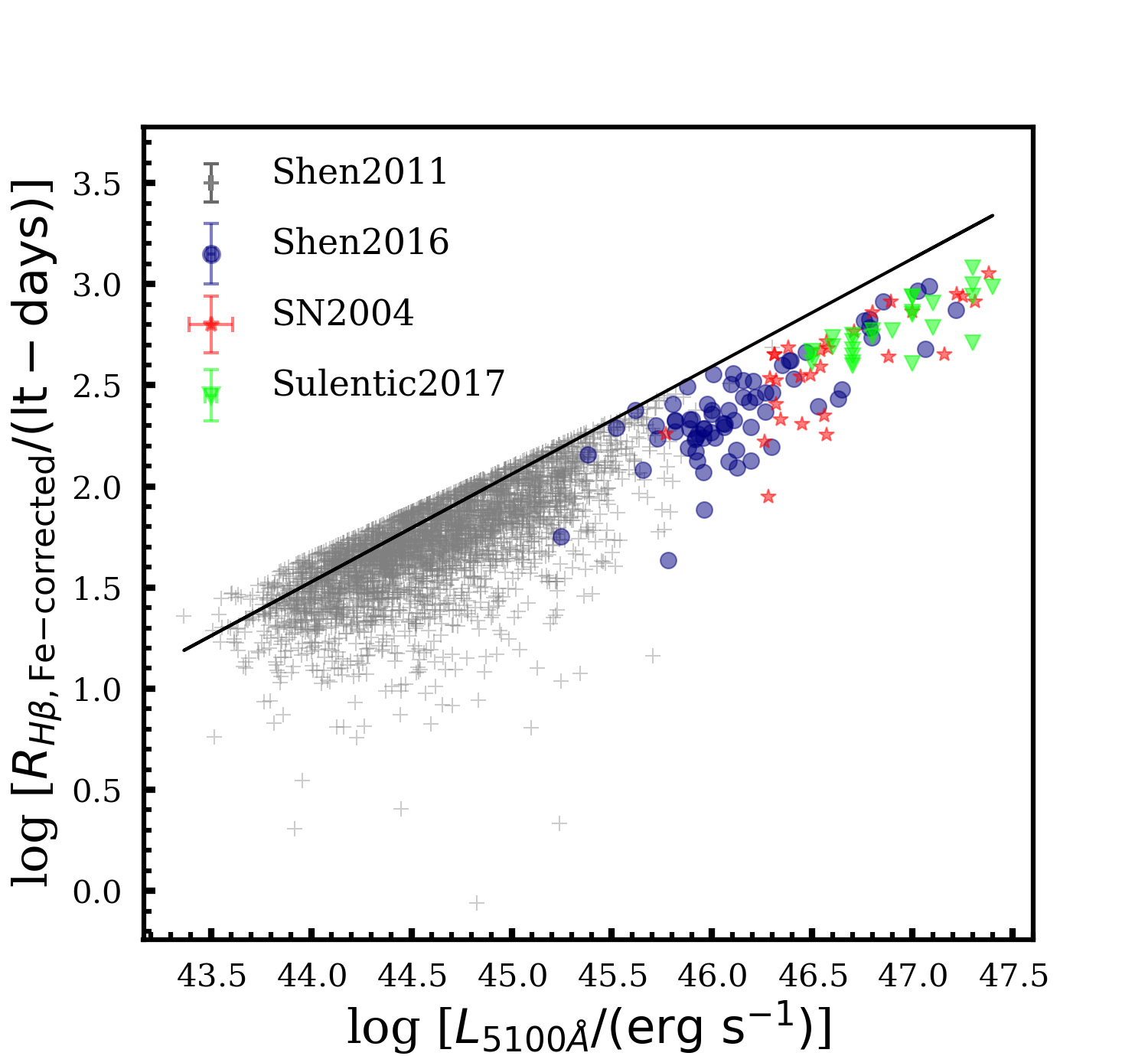}
   \caption{Fe-corrected $R_{\rm BLR}$ using equation 3 vs 5100$\angstrom$ luminosity for our quasar samples. The black solid line represents the canonical R-L relation. It shows that the Du \& Wang (2019) R-L relation predicts a smaller BLR radius than the canonical R-L relation for higher accretion rate quasars marked with strong $\mathcal{R}_{\rm Fe}$.}
   \label{fig:Figure1b}
\end{figure}
\section{Results}

Figure \ref{fig:Figure1b} shows the Fe-corrected R-L relationship (equation 3) for all the samples. For the luminous high-$z$ quasars with strong $\mathcal{R}_{\rm Fe}$, the \cite{Du_2019} R-L relation gives a smaller $R_{\rm BLR}$ as compared to the canonical R-L relationship (equation \ref{eq:eq2}). The same effect emerges in low-z quasars with strong $\mathcal{R}_{\rm Fe}$. For low-$z$ quasars with  $0\leq \mathcal{R}_{\rm Fe}\leq 0.42$, the $\mathcal{R}_{\rm Fe}$ correction is small, and the Du \& Wang (2019) R-L relation gives a larger $R_{\rm BLR}$ owing to the difference in zero-point and coefficient of luminosity ($\alpha = 1.65$, $\beta = 0.45$ )  compared to the canonical R-L relationship ($K=1.527$, $\mu = 0.533$). These differences in the predicted $R_{\rm BLR}$ from the two R-L relationships propagate to their black hole mass estimates and consequently to the accretion rates.

\subsection{The effects of the Fe correction on mass and accretion rate}
\begin{figure}
   \includegraphics[trim=0.cm 0.cm 0.9cm 0.cm,clip,width=1.\columnwidth]{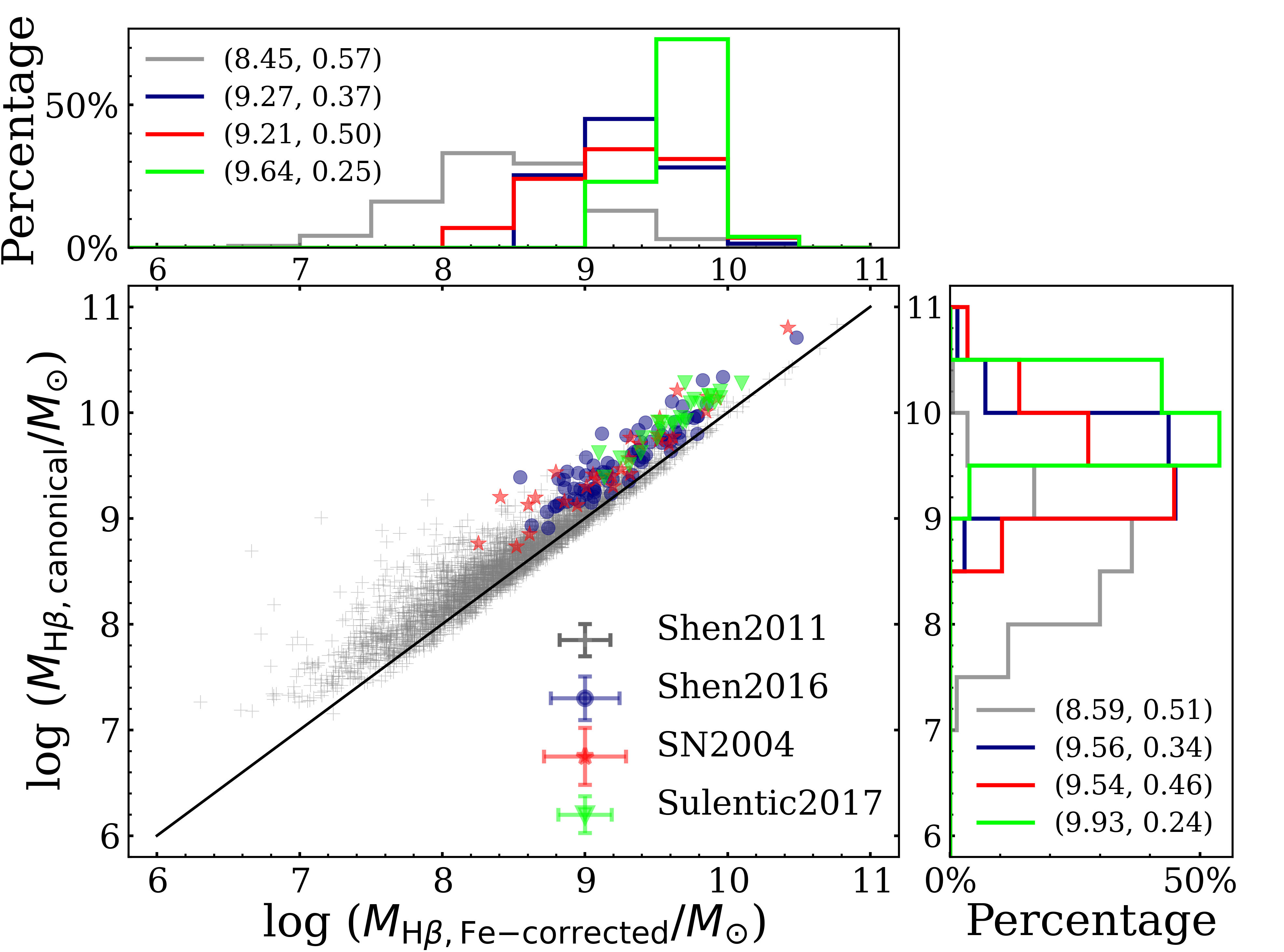}
   \caption{Black hole mass calculated using the canonical (y-axis) and Fe-corrected (x-axis) R-L relation. The histograms on top and right show the percentage distributions (i.e., the fraction of targets in each bin $\times$ 100)} of the Fe-corrected and canonical mass, respectively. The numbers in brackets give the mean and standard deviation of each sample. The plot shows that the black hole mass is systematically overestimated using the canonical R-L relation.
   \label{fig:Figure1}
\end{figure}

\begin{figure*}
   \includegraphics[trim=0.1cm 0.1cm 0.cm 0.1cm,clip,width=2\columnwidth]{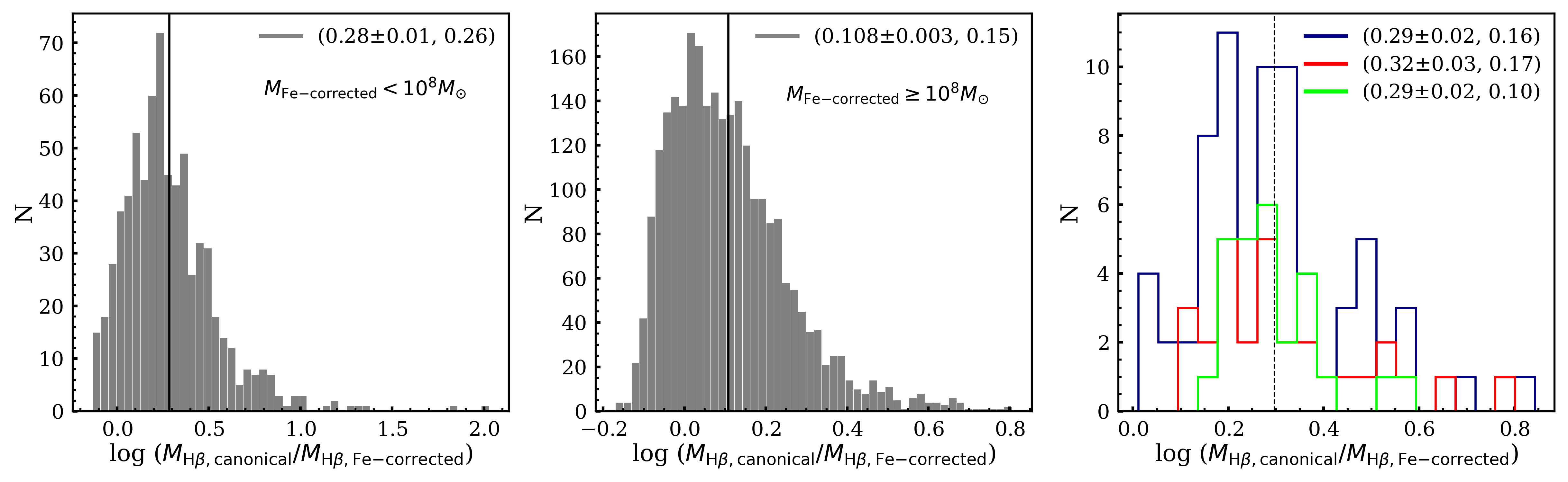}
   \caption{Left: Histogram of the distribution of mass ratios for the low-$z$ Shen2011 quasars with mass less than $10^8 M_{\sun}$. The mean $\pm$ the standard error of the mean, and the standard deviation for each sample are given in the parentheses. The mean ratio for this low-mass sub-sample is 0.28$\pm$0.01, represented by the black solid line, and the standard deviation is 0.26, as stated in parentheses. Middle: Same as left figure but for the Shen2011 quasars with mass $\geq 10^8 M_{\sun}$. The mean ratio for the high-mass sub-sample is 0.108$\pm$0.003, represented by the black solid line, and the standard deviation is 0.15, as given in parentheses. Right: Histogram of the distribution of mass ratio  for the high-$z$ samples colour-coded for each sample same as Figure 3. The individual mean and standard deviation are given in the parentheses. The mean ratio for the high-$z$ samples is 0.30$\pm$0.01, represented by the black dashed line, and the standard deviation is 0.18.}
   \label{fig:Figure2}
\end{figure*}

\begin{figure}
\includegraphics[trim=0.cm 0.4cm 0.9cm 0.0cm,clip,width=\columnwidth]{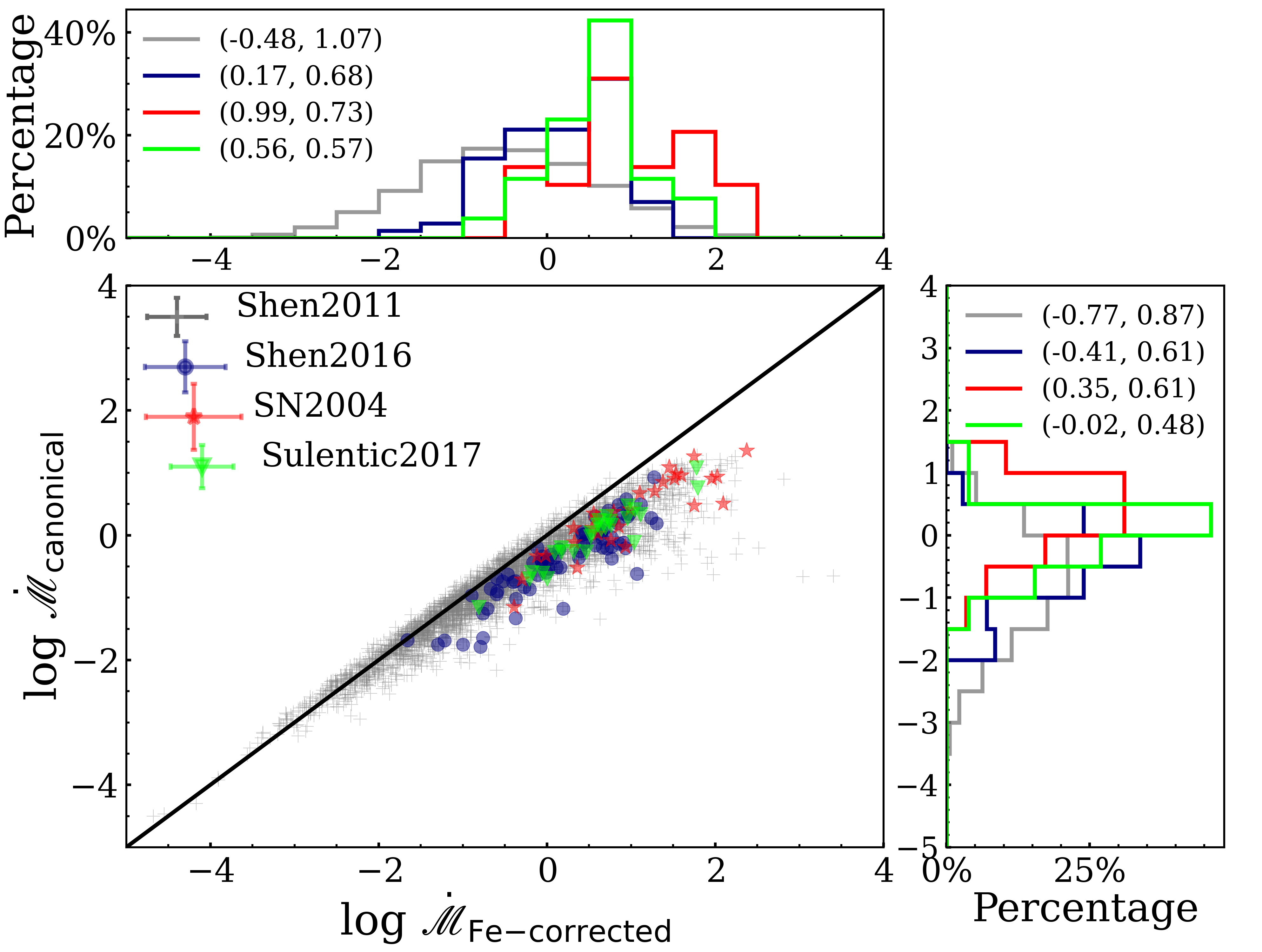}
\caption{Canonical vs Fe-corrected accretion rate calculated using equation \ref{eq:eq3}. It shows that the accretion rate is systematically underestimated when the canonical R-L relationship based black hole mass is used. The histograms on the right and top show the percentage distribution (i.e., the fraction of targets in each bin $\times$ 100) of canonical and Fe-corrected accretion rate, respectively. They show that the mean accretion rate for each sample increases when the Fe-corrected R-L relationship based mass is used.}
\label{fig:Figure4}
\end{figure}

Figure \ref{fig:Figure1} compares the Fe-corrected black hole masses against the canonical masses. The majority of the points ($~80\%$) lie above the 1:1 line, indicating a systematic overestimation of black hole mass when using the canonical R-L relation. Non-parametric tests such as Kolmogorov–Smirnov (K-S) test (p=1.84e-22) and the Wilcoxon rank sum test (two-sided p=6.27e-25) indicate that the differences in values and overall distributions between the two sets of masses significantly differ, as expected, since $\mathcal{R}_{\rm Fe}$ values are significantly above zero in significant fractions of the quasar population. It should be noted that the K-S and Wilcoxon rank sum tests do not take into account the measurement errors. We will discuss the $\mathcal{R}_{\rm Fe}$ distribution for our samples in Section 5.1.

The low-$z$ \cite{Shen_2011} quasars have a very large range in black-hole masses, $10^6< M_{\rm BH}\left (M_{\sun}\right)<10^{11}$. The higher redshift samples have a smaller range of black hole masses, $10^{8}< M_{\rm BH}\left (M_{\sun}\right)<10^{10.5}$. Figure \ref{fig:Figure1} also shows that the deviation from the 1:1 line becomes pronounced moving to the very lowest mass quasars, $M_{\rm BH} < 10^8 M_{\sun}$. These low-mass quasars have larger $\mathcal{R}_{\rm Fe}$ and higher accretion rates (see Figures \ref{fig:Figure10} \& \ref{fig:Figure11} in the Section 5.1) resulting in smaller Fe-corrected black hole masses. Therefore, in the low-$z$ quasars, the Fe-correction is most important for the less massive, highly accreting objects. The subplots on the right and top show the percentage of the canonical and Fe-corrected mass, respectively, for each sample and their statistical means and standard deviations. The mean of each sample shifts to a lower $M_{\rm BH}$ when using the \citeauthor{Du_2019} R-L relation. Note that \cite{Shen_2011} sample consists of a tiny population (20 out of 3309) of quasars with $10^6\leq M_{\rm BH} \left (M_{\sun}\right) \leq 10^7$ for which $f=0.7\pm0.2$ may be more appropriate \citep[][for the low mass AGNs with pseudobulges]{Ho&Kim2014}. Our choice of $f=1.5$ likely overestimates both the canonical and Fe-corrected black hole masses of these 20 quasars alike, although without impacting the ratio of the canonical to Fe-corrected mass.

To further illustrate the change in mass when the Fe correction is applied, we plot the distribution of the log of the ratio of canonical-to-Fe-corrected mass for the low-$z$ \cite{Shen_2011} sample in the left and the middle panels of Figure \ref{fig:Figure2}. The left histogram consists of quasars with $M_{\rm Fe-corrected}< 10^{8} M_{\sun}$, whereas the middle panel shows quasars with $M_{\rm Fe-corrected} \geq 10^{8} M_{\sun}$. The low-mass quasars ($< 10^{8} M_{\sun}$), on average, have a much larger overestimation, a factor of $\sim 2$, compared to a factor of $\sim 1.3$ for the high-mass ($\geq 10^{8} M_{\sun}$) quasars in the low-$z$ sample. The low-mass quasars are the ones with the strongest $\mathcal{R}_{\rm Fe}$ (see Figure \ref{fig:Figure11}). For the low-z quasars with weaker $\mathcal{R}_{\rm Fe}$, using the \cite{Du_2019} R-L relation slightly overestimates the black hole mass of $\sim$22\% of the sample compared to the canonical, owing to small differences in intercept and coefficient of luminosity.

The right histogram of Figure \ref{fig:Figure2} illustrates the change in mass for samples from Table 1, which excludes \cite{Shen_2011}.  On average, the black hole masses of these high-$z$ samples decrease by about 0.3 dex or a factor of 2, as depicted by the dotted black line. The difference is up to a factor of 2 for 60\% of the sample and a factor of 2-4.7 for 37\%. The mean difference in mass in \cite{Shen2016} is a factor of $\sim$2, in SN 2004 is a factor of $\sim$2.1, and in \cite{Sulentic2017} is a factor of $\sim$2.

\begin{figure}
\includegraphics[trim=1.5cm 1.0cm 1.5cm 1.2cm,clip,width=0.8\columnwidth]{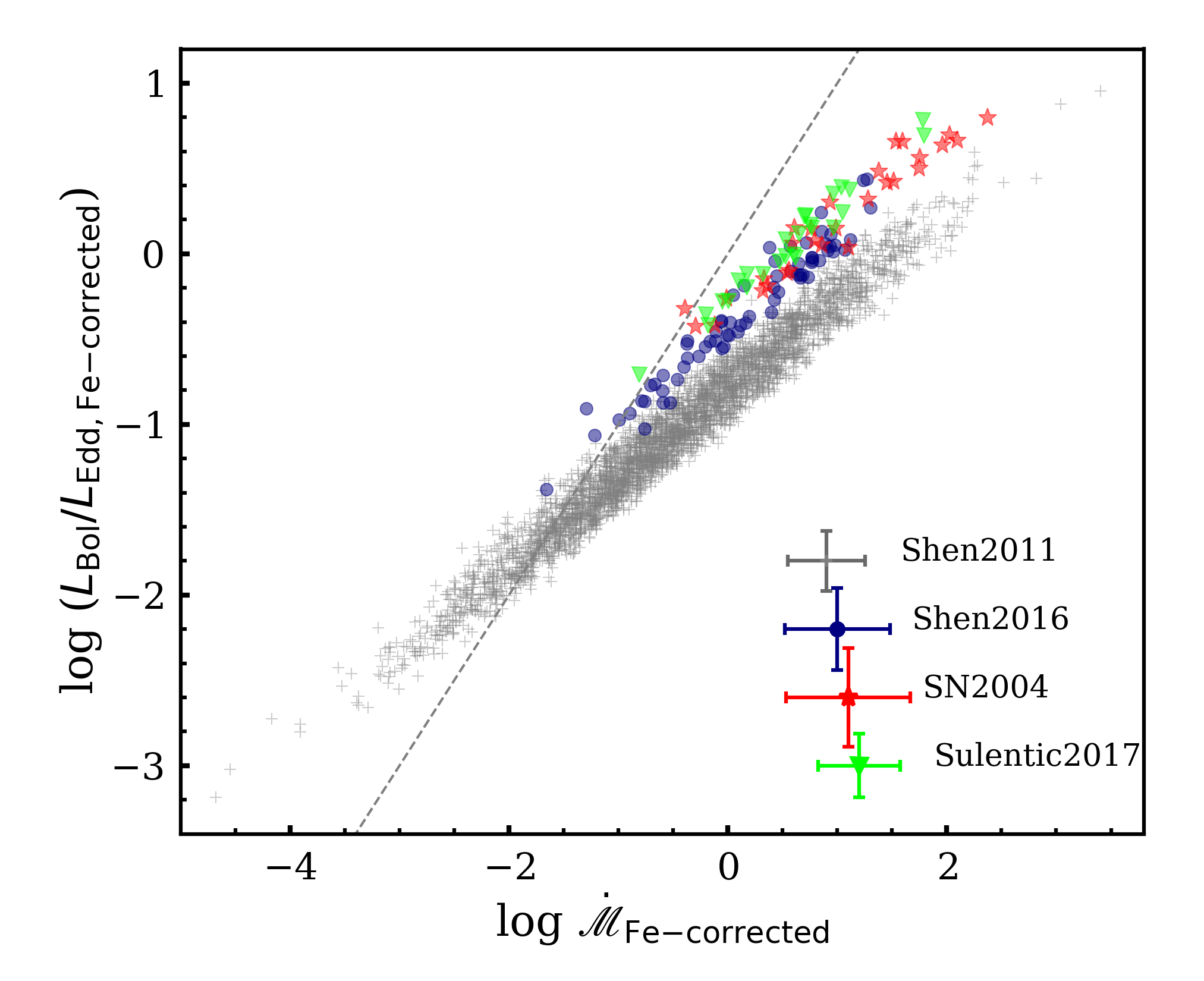}
\caption{Corrected Eddington ratio vs accretion rate parameter for all samples. The grey dashed line represents the 1:1 relation for reference.}
\label{fig:Figure5}
\end{figure}

Next, we calculated the dimensionless accretion rate parameter defined in equation \ref{eq:eq4} using both canonical and Fe-corrected black hole mass. The comparison between canonical and Fe-corrected $\dot{\mathscr{M}}$ (Figure \ref{fig:Figure4}) demonstrates that the accretion rates are underestimated, in agreement with the inverse relationship between $\dot{\mathscr{M}}$ and the black hole mass. The distribution in Figure \ref{fig:Figure4} shows that the mean $\dot{\mathscr{M}}$ for each sample is larger for the Fe-corrected black hole mass. For low-z quasars, i.e., the Shen et al. (2011) sample, the logarithm of $\dot{\mathscr{M}}_{\rm Fe-corrected}$ has a mean $\pm$ the standard error of the mean\footnote{Note that the standard error of the mean quoted in the figures and text is purely statistical in nature (ratio of standard deviation and square root of sample size) and does not take into account the measurement uncertainties.} of -0.48$\pm$0.02, and it ranges from $-4.68$ to $3.40$. Whereas the high-z quasars (samples other than Shen et al. 2011) have systematically higher log $\dot{\mathscr{M}}_{\rm Fe-corrected}$ values with a mean of 0.44$\pm$ 0.07 and range from $-1.66$ to $2.37$. Figure \ref{fig:Figure5} compares $\dot{\mathscr{M}}$ with the traditionally used Eddington ratio, both calculated using the Fe-corrected black hole mass. The log of the Fe-corrected Eddington ratio for the low-z quasars ranges from $-$3.19 to 0.95 and has a mean value of $-$1.04$\pm$0.01, whereas the high-z quasars have a higher mean value of log $L_{\rm Bol}/L_{\rm Edd, Fe-corrected} = -$0.13$\pm$ 0.04 and range from $-1.38$ to $0.80$. 

\begin{figure}
\begin{center}
\includegraphics[trim=1.cm 1.1cm 0.5cm 0.1cm,clip,width=0.8\columnwidth]{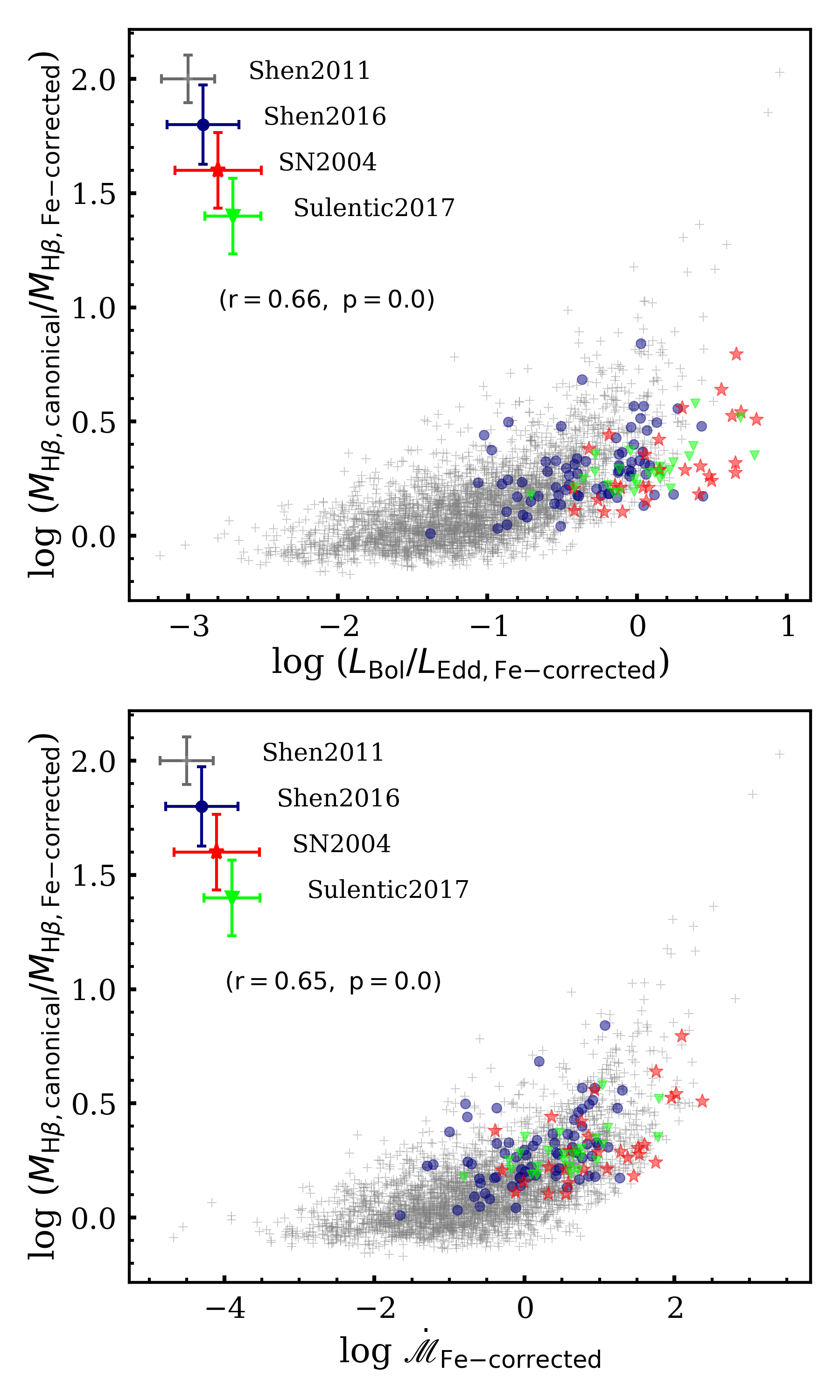}
\caption{Mass ratio vs Fe-corrected Eddington ratio (top panel) and dimensionless accretion rate (bottom panel). Change in mass shows a strong correlation with the accretion rate parameters. The Pearson r coefficients for these correlations are given in parentheses.  }
\label{fig:Figure6a}
\end{center}
\end{figure}

As mentioned earlier, the canonical and Fe-corrected black hole mass differ primarily due to the $\mathcal{R}_{\rm Fe}$ factor, an accretion rate indicator. Therefore, the plot of the change in the black hole mass or the mass ratio exhibits the expected strong correlation with the two accretion rate parameters, $L_{\rm Bol}/L_{\rm Edd}$ \& $\dot{\mathscr{M}}$ (Figure \ref{fig:Figure6a}).  We visually inspected the SDSS spectra of the \citet{Shen_2011} quasars with the log of the mass ratio $\geq1$, which appear as outliers in Figure \ref{fig:Figure6a}. Their spectra are consistent with the EV1 trend of strong relative strength of Fe {\sc ii} (see Appendix \ref{app:app2}). 

\begin{figure}
\begin{center}
\includegraphics[trim=0.8cm 0.5cm 0.5cm 0.5cm,clip,width=\columnwidth]{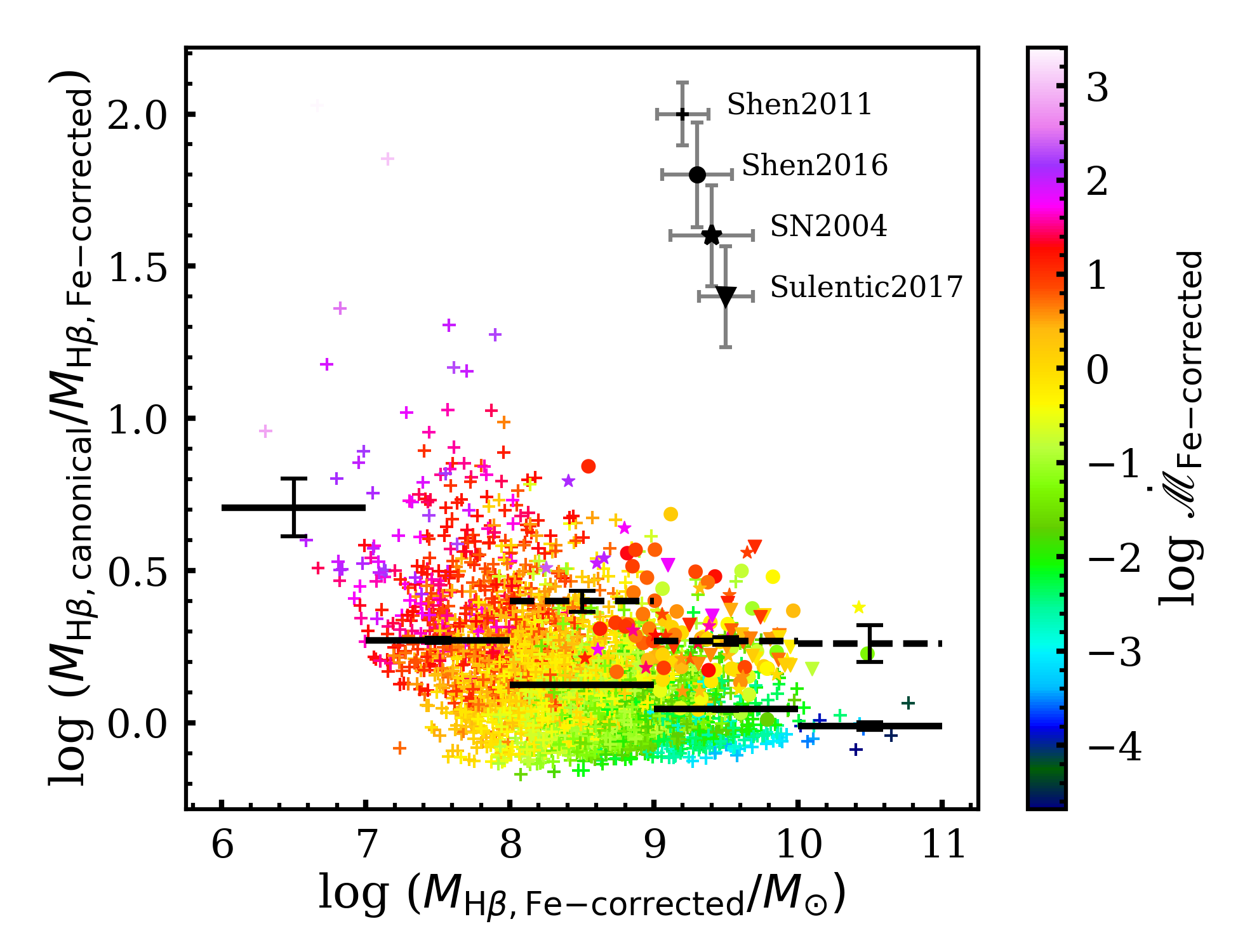}
\caption{Mass ratio vs Fe-corrected black hole mass colour-coded by accretion rate. Black solid and dashed lines represent the mean mass ratio in each mass bin for the low-$z$ sample and the high-$z$ samples, respectively.  The black error bars represent the standard error of the mean ratio in each mass bin. The grey error bars on the top of each sample represent the typical measurement error in the mass ratio. For $M_{\rm H\beta, Fe-corrected}>10^8 M_{\sun}$, the high-z samples have a mean mass ratio greater than the low-z sample.}
\label{fig:Figure6}
\end{center}
\end{figure}
\begin{figure}
\begin{center}
\includegraphics[trim=0.5cm 0.2cm 2.cm 3.0cm,clip,width=\columnwidth]{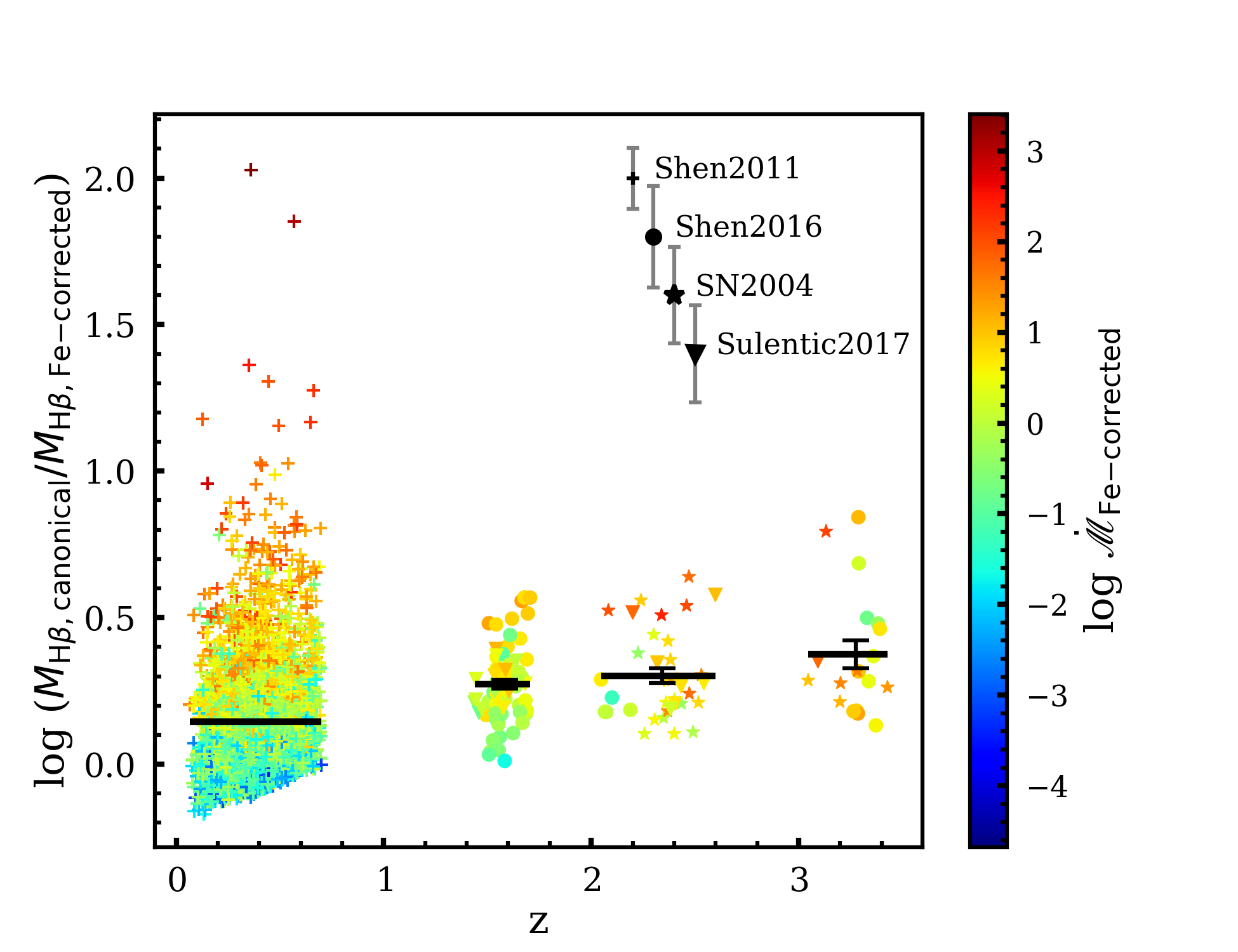}
\caption{Mass ratio between canonical and Fe-corrected black hole mass vs redshift, colour-coded by accretion rate. Black solid lines show the mean mass ratio within the redshift range presented by the length of the line. Black error bars shows the standard error of the mean mass ratio. The plot shows that the difference between the canonical and Fe-corrected mass increases as the redshift increases. The mean mass ratio in redshift bins $2\leq z\leq3$ and $z>$3 differ at the  $\sim$1.1$\sigma$ level. On average, the high-$z$ quasars ($z>1$) have higher accretion rates.}
\label{fig:Figure7}
\end{center}
\end{figure}

\begin{figure}
\begin{center}
\includegraphics[trim=0.0cm 0.2cm 2.cm 3.0cm,clip,width=\columnwidth]{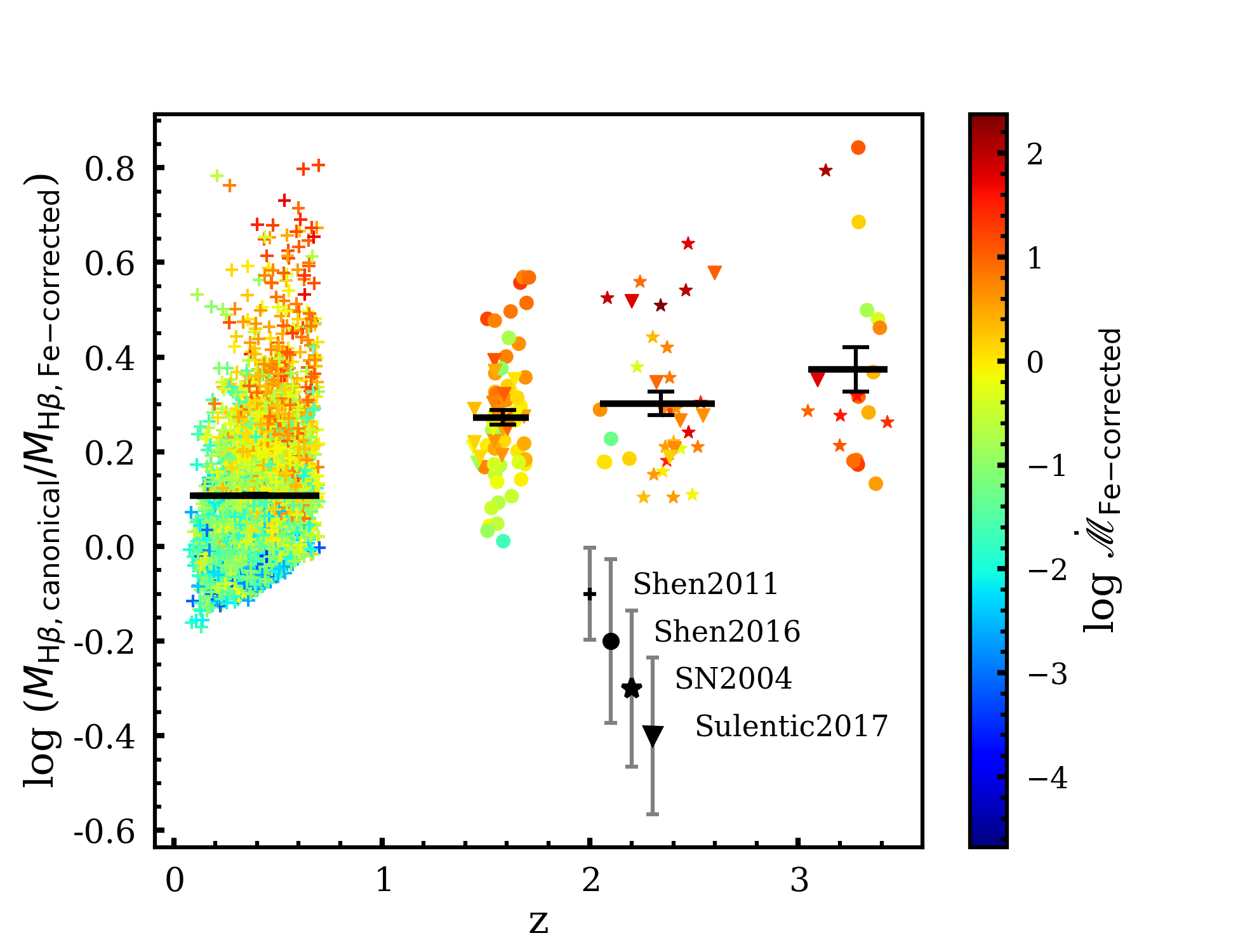}
\caption{Same as Figure 9 with only $M_{\rm H\beta, Fe-corrected}>10^8 M_{\sun}$ targets with $z<0.7$.}
\label{fig:Figure7b}
\end{center}
\end{figure}

Figure \ref{fig:Figure6} displays the mass ratio plotted against the Fe-corrected mass colour-coded by the Fe-corrected accretion rate. It shows the mean mass ratio in each mass bin for low and high-$z$ samples separately. The mean mass ratio and the standard error of the mean in each mass bin for the low-$z$ sample are 0.71$\pm$0.09 dex in $10^6\leq M_{\rm BH}\left (M_{\sun}\right)<10^7$, 0.27$\pm$0.01 dex in $10^7\leq M_{\rm BH}\left (M_{\sun}\right)<10^8$, 0.125$\pm$0.003 dex in $10^8\leq M_{\rm BH}\left (M_{\sun}\right)<10^9$, 0.045$\pm$0.004 dex in $10^9\leq M_{\rm BH}\left (M_{\sun}\right)<10^{10}$ and -0.01$\pm$ 0.01 dex in $M_{\rm BH}\geq 10^{10} M_{\sun}$. The mean mass ratio and the standard error of mean in each mass bin for the high-$z$ samples are 0.40$\pm$0.03 dex in $10^8\leq M_{\rm BH}\left (M_{\sun}\right)<10^9$, 0.27$\pm$0.01 dex in $10^9\leq M_{\rm BH}\left (M_{\sun}\right)<10^{10}$ and 0.26$\pm$0.06 dex in $M_{\rm BH}\geq 10^{11} \left (M_{\sun}\right)$. Figure \ref{fig:Figure6} demonstrates that the lower mass quasars ($M_{\rm BH} < 10^8 M_{\sun}$) of the low-$z$ \cite{Shen_2011} sample have a higher accretion rate, hence, larger mass correction. For $M_{\rm BH} \geq 10^8 M_{\sun}$, the mass correction is, on average, larger in the high-z samples compared to the low-z sample.

\begin{figure*}
\begin{center}
\includegraphics[trim=8.5cm 0.0cm 0.cm 0.5cm,clip,width=2.2\columnwidth]{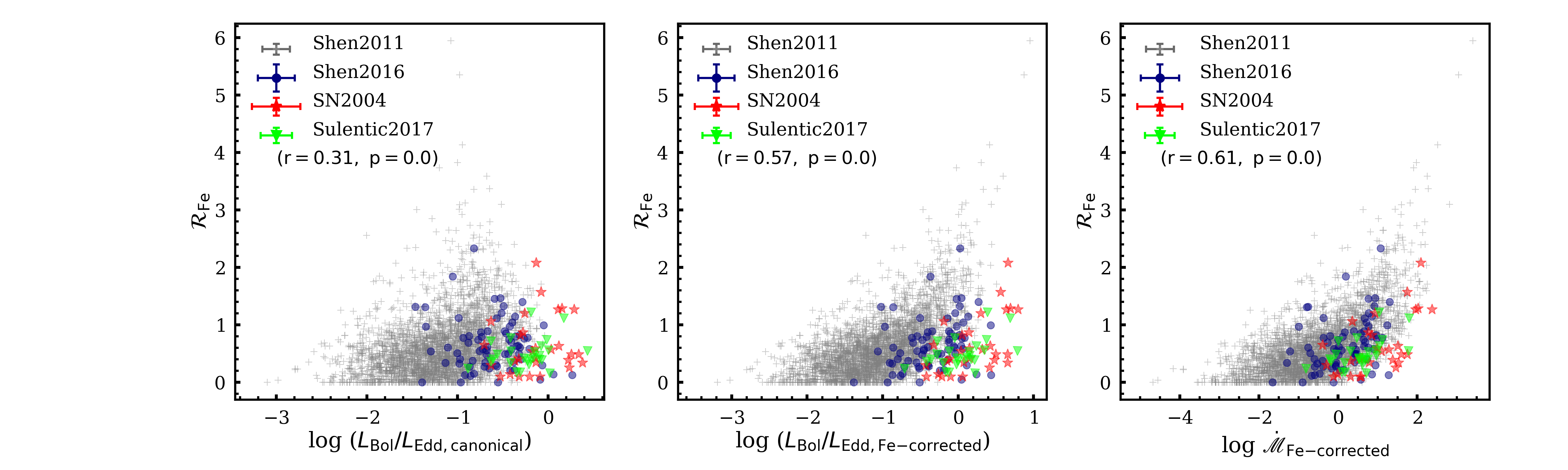}
\caption{$\mathcal{R}_{\rm Fe}$ vs canonical Eddington ratio (left), Fe-corrected Eddington ratio (middle) and Fe-corrected $\dot{\mathscr{M}}$ (right). The Pearson correlation coefficients $r$ and $p$ are given in parentheses.}
\label{fig:Figure8}
\end{center}
\end{figure*}

Plotting the change in black hole mass against redshift, colour-coded by accretion rate, Figure \ref{fig:Figure7} shows that, on average, the overestimation of mass increases as the redshift increases. The mean mass differences is 0.145$\pm$0.003 dex for $z<1$, 0.27$\pm$0.01 dex for $1<z<2$, 0.30$\pm$0.02 dex for $2<z<3$ and 0.37$\pm$0.05 dex for $z>3$. Figure \ref{fig:Figure7} also demonstrates that the high-$z$ ($z>1$) quasars on average have higher accretion rate black holes as compared to the low-$z$ quasars, which include a large fraction of high mass but low-accretion rate objects (also see Figure  \ref{fig:Figure10}). It also shows that for $z>1$, the fraction of high accretion rate quasars increases as redshift increases and their masses are overestimated by a factor of two to several using the canonical R-L relationship. Figures \ref{fig:Figure6} \& \ref{fig:Figure7} show that the high-$z$ quasars have larger masses and mass correction; they also have larger accretion rates on average (see Figure \ref{fig:Figure6a}) -- the high-$z$ quasars lack quasars with very weak Fe {\sc ii} that are present in large numbers in the low-redshift Shen et al. (2011) sample (see Figure \ref{fig:Figure11}).  This is likely a selection effect due to the difficulty of observing lower luminosity quasars at high-$z$.

Figure \ref{fig:Figure7b} is qualitatively the same as Figure \ref{fig:Figure7} but includes only $M_{\rm BH} > 10^8 M_{\odot}$ quasars from the low redshift \cite{Shen_2011} sample. This eliminates 692 quasars with strong $\mathcal{R}_{\rm Fe}$ and high accretion rate, giving a mean mass difference of 0.108$\pm$0.003 dex for this low-z sub-sample. Figure \ref{fig:Figure7b} demonstrates the stark difference the Fe-correction makes on the quasars with comparable black hole masses in the low and high-redshift samples.

\subsection{A stronger correlation between $\mathcal{R}_{\rm Fe}$  and accretion rate parameters after using Fe-corrected mass}
The dominant trend of decreasing EW[O{\sc iii}] with the increasing $\mathcal{R}_{\rm Fe}$ is known as Eigenvector 1 (EV1) \citep{BG1992}. It represents a correlation space in which many quasars properties correlate with the optical Fe {\sc ii} strength \citep{BG1992, Boroson2002}. EV1 is primarily governed by the black hole accretion process parameterized by the Eddington ratio \citep{BG1992, Boroson2002, Yuan+2003, Marziani+2001, Shen_Ho2014, Sun&Shen2015}. The correlation between mass ratio and accretion rate parameters seen in Figure \ref{fig:Figure6a} is largely due to this EV1 dependence.

\citet[Figure 1]{Du_etal2016} show the correlation between $\mathcal{R}_{\rm Fe}$ and $L_{\rm Bol}/L_{\rm Edd}$ (also, $\dot{\mathscr{M}}$) for the SDSS DR5 sample of \cite{Hu+2008} and RM AGNs. We plot versions of this correlation in Figure \ref{fig:Figure8} for our low and high-$z$ samples. The left panel of Figure  \ref{fig:Figure8} demonstrates that using an underestimated Eddington ratio based on the canonical black hole mass ($L_{\rm Bol}/L_{\rm Edd} \propto M_{\rm BH}^{-1}$) gives a Pearson $r$ correlation coefficient of only 0.31. The Fe-corrected black hole mass more accurately determines a higher Eddington ratio for the strong $\mathcal{R}_{\rm Fe}$ emitters, improving the correlation to $r=0.57$ (Figure \ref{fig:Figure8}, middle panel). The right panel of Figure \ref{fig:Figure8} shows the strongest correlation, $r=0.61$, between $\mathcal{R}_{\rm Fe}$ and Fe-corrected $\dot{\mathscr{M}}$. We recognize that the much larger correlation coefficient is in part an effect of self-correlation induced by the addition of $\mathcal{R}_{\rm Fe}$ factor in the \cite{Du_2019} R-L relationship to estimate the Fe-corrected mass and consequently the Fe-corrected accretion rate parameters.

\section {Discussion}

\subsection{Selection biases and overall distribution of bright quasar properties}
It is worth noting that the samples we use are not comprehensive and have selection effects. These samples, however, are likely representative of bright quasars at low and high redshifts and demonstrate the importance of the Fe-correction on the black hole mass estimates of highly accreting AGNs, especially at high-redshift.

\begin{figure*}
   \includegraphics[trim=0.cm 0.cm 0.5cm 2.cm,clip,width=\columnwidth]{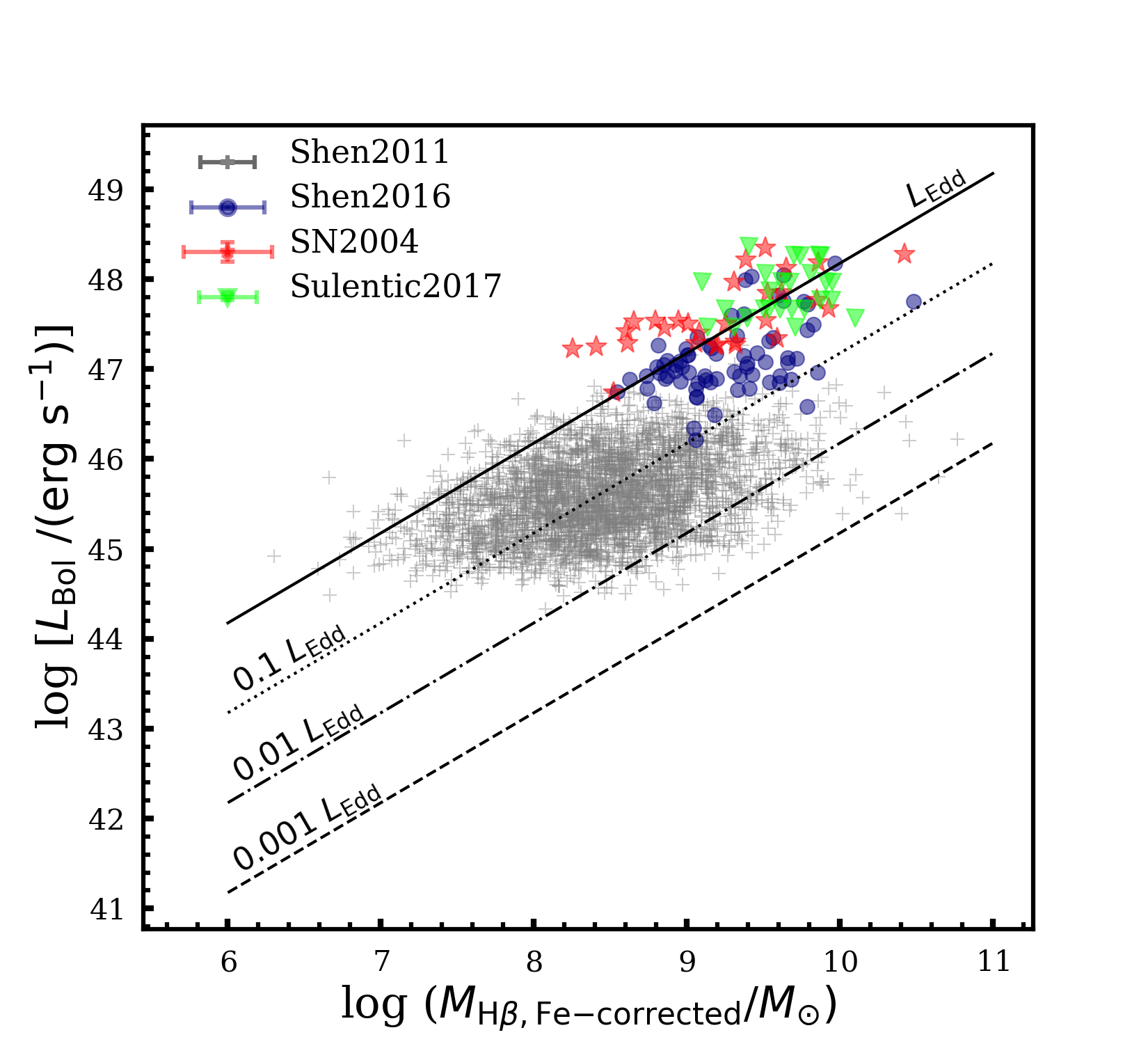}
   \caption{Bolometric luminosity vs Fe-corrected black hole mass for all samples. The black solid line represents the $L_{\rm Bol}=L_{\rm Edd}$, dotted line represents $L_{\rm Bol}=0.1 L_{\rm Edd}$, dotted-dashed line represents $L_{\rm Bol} = 0.01 L_{\rm Edd}$ and dashed line represents $L_{\rm Bol} = 0.001 L_{\rm Edd}$. }
   \label{fig:Figure9}
   \centering
   \begin{subfigure}{0.5\textwidth}
     \centering
     \includegraphics[trim=1.cm .5cm 1.5cm 2.cm,clip,width=\columnwidth]{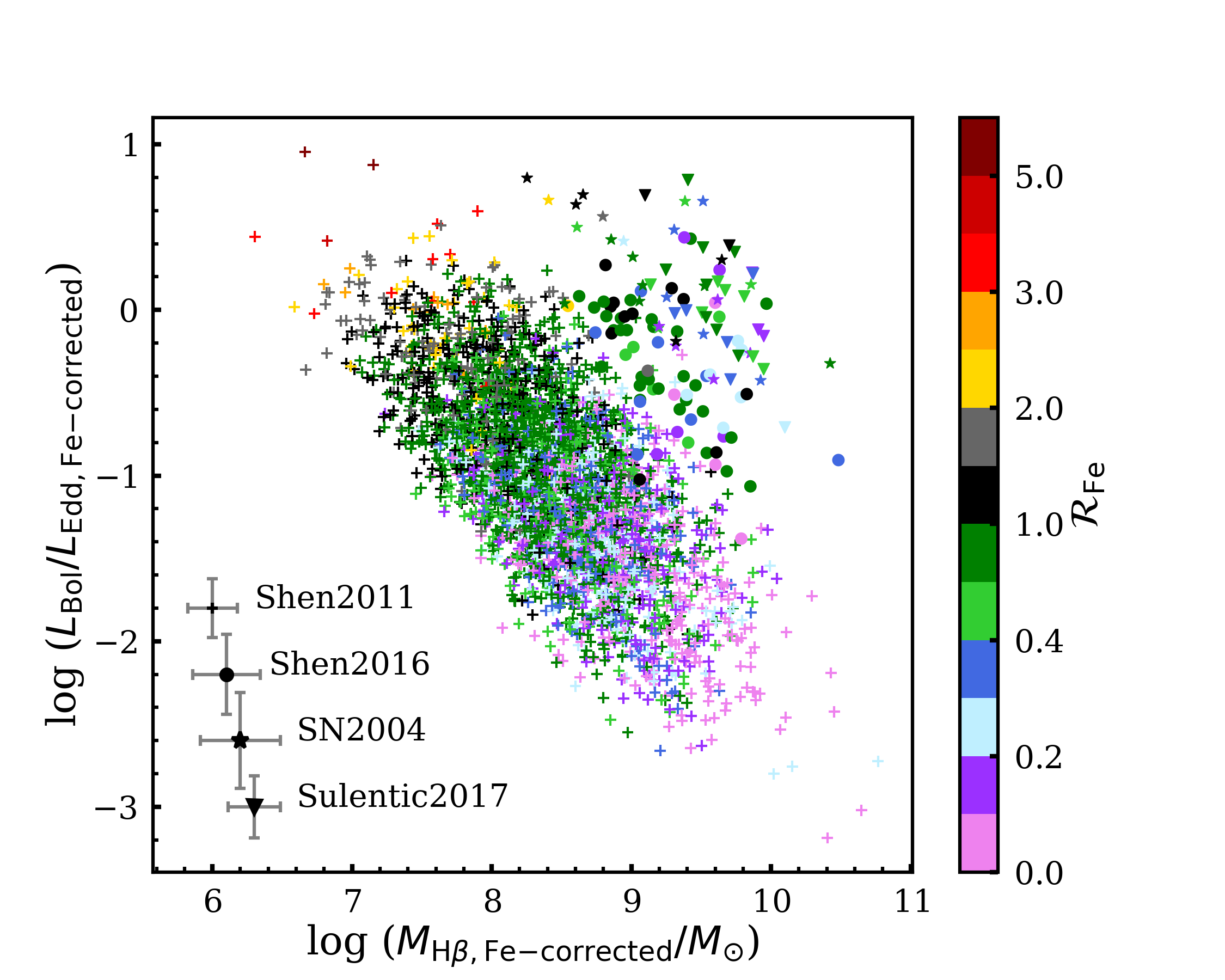}
   \end{subfigure}%
   \begin{subfigure}{0.5\textwidth}
     \centering
     \includegraphics[trim=1.cm .5cm 1.5cm 2.cm,clip,width=\columnwidth]{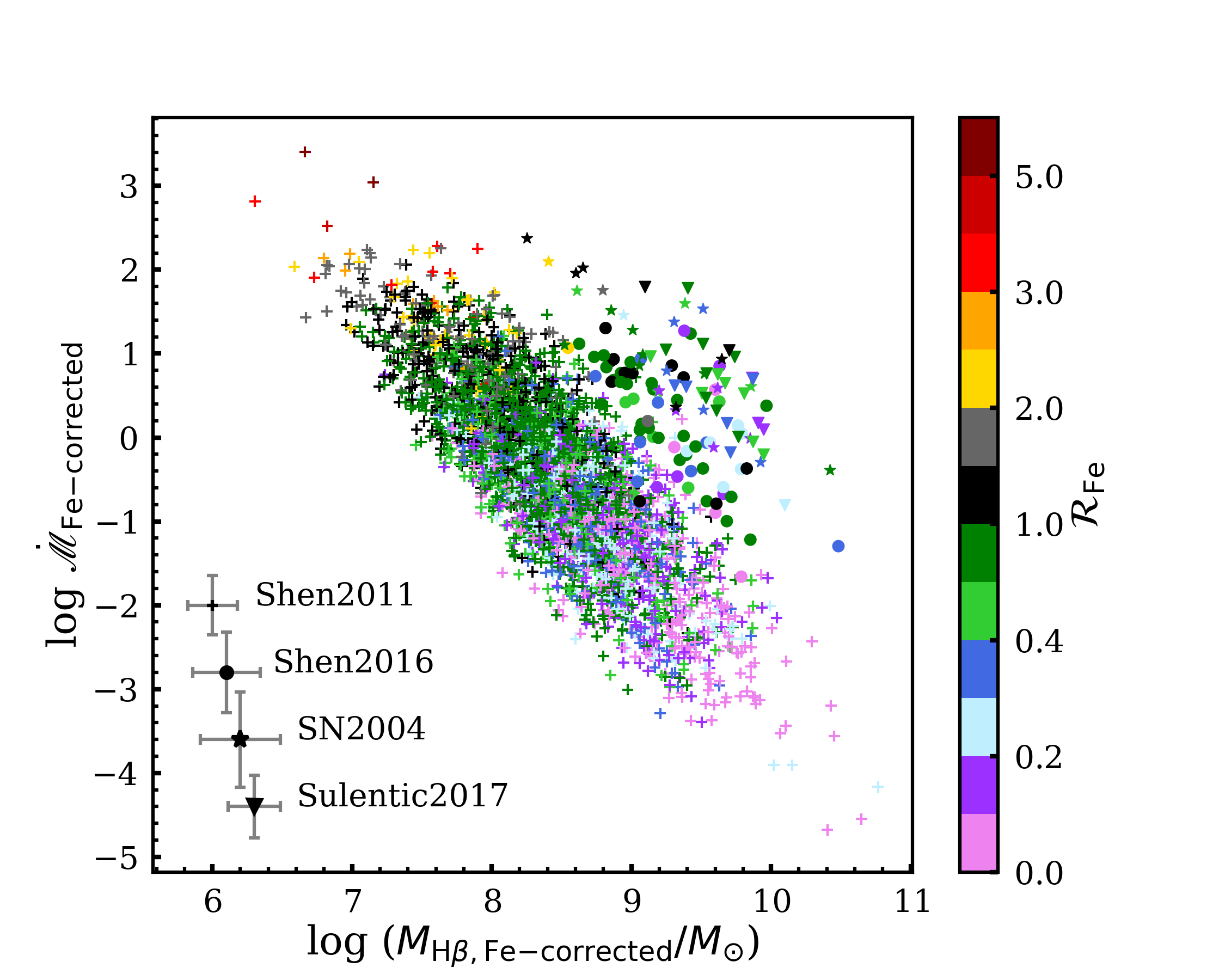}
   \end{subfigure}%
     \caption{Fe-corrected mass vs accretion rate parameter colour-coded by $\mathcal{R}_{\rm Fe}$.}
     \label{fig:Figure10}
\end{figure*}

The luminosities and redshifts of the quasars we use depend on various selection effects, such as adopted signal-to-noise ratio cuts and the wavelength of red-shifted H$\beta$ and if it falls in a spectral window observable from the ground, as well as the properties of the quasar population itself.
For instance, there are no extremely luminous quasars at low-redshift, as 
only the lower mass objects are actively accreting, whereas at high
redshift it is the most massive systems that are actively forming,
a phenomenon known as ``downsizing'' \citep[][]{Heckman+2004, Hasinger+2005}.
To show these and other effects, we plot the mass-luminosity plane for all our samples in Figure \ref{fig:Figure9}.  We see that there are quasars accreting at or slightly above the Eddington luminosity in both the low and high-redshift samples, but they are of intrinsically different masses and luminosity.  Furthermore, it
is only a small fraction of the low-redshift quasars that show such high 
accretion rates, as compared to the high-redshift samples.  There are de facto luminosity cuts for the lowest mass black holes  ($<10^8 M_{\sun}$), which must be accreting at Eddington ratios greater than $\sim$0.1 to be luminous enough to be included in the sample.  The near-IR spectral surveys naturally chose the brightest objects known at high-redshift, which correspond to the most luminous and highest accretion rate quasars with the largest black hole masses.  There is also a de facto accretion rate limit below $\sim$0.01 of the Eddington ratio, where active galaxies stop displaying broad emission lines \citep[e.g.,][and references within]{Guolo+2021}.

\begin{figure}
   \includegraphics[trim=0.cm 0.cm 0.5cm 2.cm,clip, width=\columnwidth]{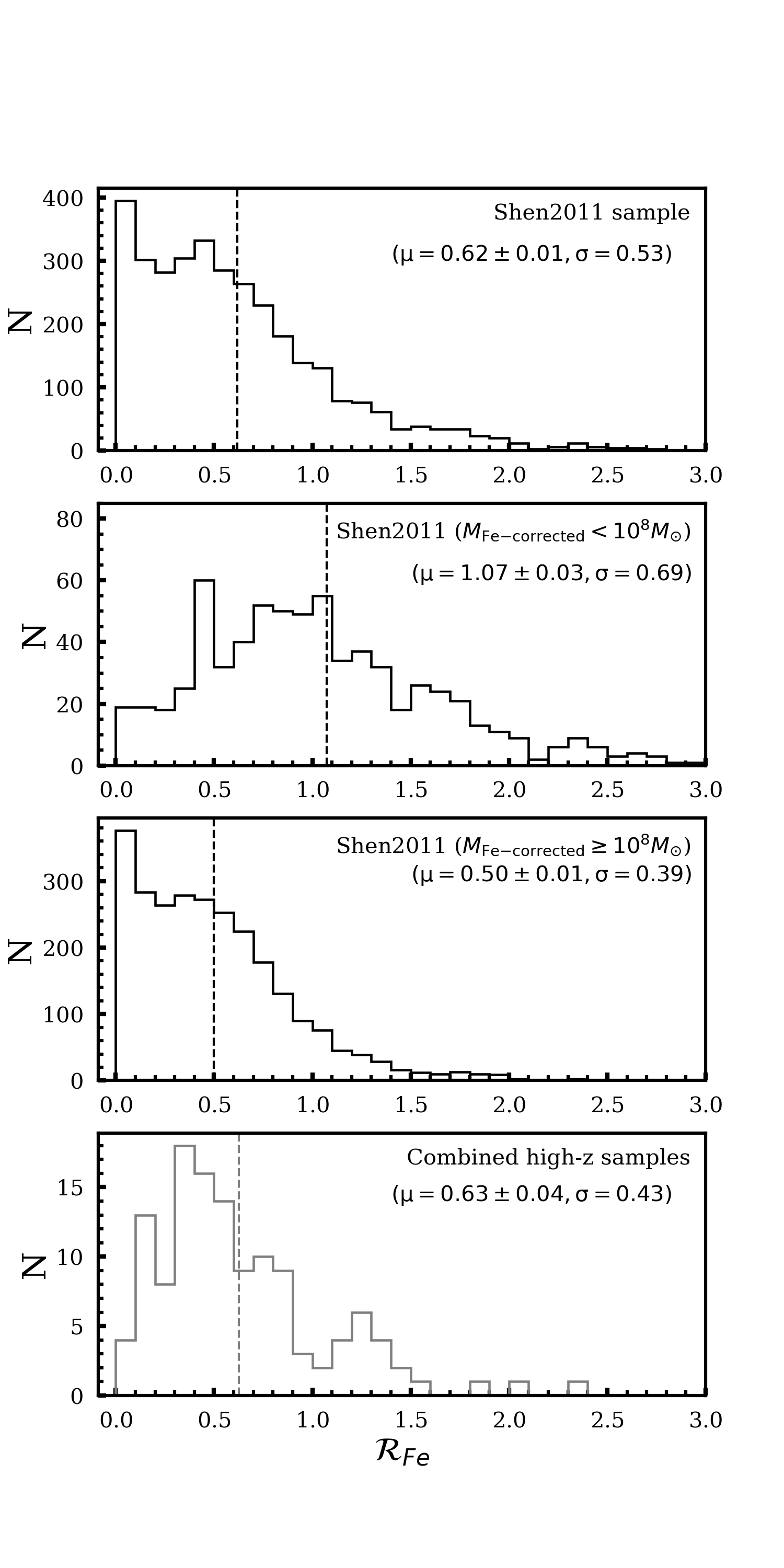}
   \caption{Distribution of $\mathcal{R}_{\rm Fe}$ in low-redshift sub-samples of Shen et al. 2011 (top three panels) and all the high-redshift samples combined together (bottom panel). The plot excludes targets with $\mathcal{R}_{\rm Fe}>3$. }
   \label{fig:Figure11}
\end{figure}

Figure \ref{fig:Figure10} plots our Fe-corrected black hole masses against our accretion rate indicators, and features a colour scheme to identify the values of $\mathcal{R}_{\rm Fe}$ across the plane.  The super-Eddington accreting quasars in both low and high-redshift samples in general display the largest $\mathcal{R}_{\rm Fe}$ measurements.  The quasars with large black hole masses and the lowest accretion rates in general display the smallest $\mathcal{R}_{\rm Fe}$ measurements. We show the distribution of $\mathcal{R}_{\rm Fe}$ in the low-$z$ and the combined high-$z$ samples in Figure \ref{fig:Figure11}. To illustrate the diversity in accretion rates of low-$z$ quasars, we further divided the low-z sample into low and high mass. There are 692 out of 3309 quasars in the low-$z$ sample with a mass less than $10^8 M_{\sun}$, while the remaining 79\% have {a} mass $\geq 10^8 M_{\sun}$. We excluded 11 quasars in the low-mass low-z sample (spectra shown in Appendix B) that have $\mathcal{R}_{\rm Fe}>3$. Figure \ref{fig:Figure11} demonstrates that low mass, low-$z$ quasars have higher accretion rates (Figure \ref{fig:Figure10}) and stronger $\mathcal{R}_{\rm Fe}$ (Figure \ref{fig:Figure11}). They are the low mass analogs of massive highly accreting quasars at high-$z$, albeit less massive and less luminous. On the other hand, the high-mass quasars in the low-$z$ sample have lower accretion rates and smaller $\mathcal{R}_{\rm Fe}$ measurements. The distribution of $\mathcal{R}_{\rm Fe}$ in the high-$z$ quasars shows that they consist of strong Fe {\sc ii} emitters and generally lack quasars with very small or zero $\mathcal{R}_{\rm Fe}$.

\subsection{Importance of the $\mathcal{R}_{\rm Fe}$  correction}
Two dimensionless quantities commonly used to parameterize the black hole's accretion rate are Eddington ratio ($L_{\rm Bol}/L_{\rm Edd}$) and $\dot{\mathscr{M}}$. With the increasing number of RM measurements, many studies use these quantities to correct the R-L relationship for accretion-rate dependence. \cite{Alvarez_etal2020} point out that correlations between an $R_{\rm BLR}$ offset, i.e., the difference between $R_{\rm BLR}$ from RM measurement and one estimated from R-L relationship, and accretion rate estimators suffer from self-correlation. The $R_{\rm BLR}$ offset is proportional to the ratio of $R/L^{0.5}$, $L_{\rm Bol}/L_{\rm Edd} \propto L_{5100\angstrom}/R$ and $\dot{\mathscr{M}} \propto L_{5100\angstrom}^{1.5}/R^2$. Therefore, using an independent estimate of accretion rate like $\rm \mathcal{R}_{Fe}$ is better. A strong anti-correlation is seen between the BLR size offset and $\rm \mathcal{R}_{Fe}$ for RM AGNs in the SEAMBH sample \citep{Du_etal2018, Du_2019}.  \cite{Grier_etal2017}, however, studied their H$\beta$ SDSS-RM sample and did not find this anti-correlation \citep{Alvarez_etal2020}. The H$\beta$ SDSS-RM sample has a median $L_{\rm Bol}/L_{\rm Edd} \sim 0.1$ \citep{Shen_etal2019} and lacks extremely high accretors. For low-accreting AGNs, both accretion rate and, perhaps, black hole spin, govern the BLR size \citep{Wang_etal2014, Du_etal2018}. After dividing the SDSS-RM sample into high and low accretion rate, \cite{Du_etal2018} show that the high accretion rate sub-sample follows the expected anti-correlation between the BLR size offset and $\rm \mathcal{R}_{Fe}$ (see Figure 5 of \citealt{Du_etal2018}). 

Our results demonstrate the significant impact of including $\mathcal{R}_{\rm Fe}$ in the R-L relationship on the determination of black hole mass and accretion rate parameters. Such accretion rate-based correction is crucial for luminous broad-absorption line (BAL) quasars that are known to have strong Fe {\sc ii} \citep[][ and references therein]{Turnshek1997, Boroson2002, Yuan+2003, Runnoe+2013a}. We found that the strongest Fe-emitters in general have higher accretion rate, and consequently overestimated black hole mass. Quasars with $L_{\rm Bol}/L_{\rm Edd} \geq 1$ can have black hole mass overestimated by up to an order of magnitude (as shown in Figure \ref{fig:Figure6a}).

\subsection{Some caveats for black hole mass estimation}
We made some choices, depending on the availability of data, that impacted our results. Past studies interchangeably used flux ratio and EW ratio to define  $\mathcal{R}_{\rm Fe}$. To check how much this impacts estimates of black hole mass, we evaluated the change in $\mathcal{R}_{\rm Fe}$ by using each of these two quantities, in turn, to estimate black hole masses for the \cite{Shen_2011} sample. They tabulate EWs of H$\beta$ and Fe {\sc ii}. We used the given EW and line luminosity for H$\beta$ to compute the continuum luminosity at 4861$\angstrom$. We scaled the continuum luminosity to the mean wavelength of Fe {\sc ii} i.e., 4560$\angstrom$ using the slope for the H$\beta$ region and calculated the line luminosity for Fe {\sc ii}. The mean percentage difference between $\rm \mathcal{R}_{Fe}$ measured from line luminosity ratios and EW ratios is $3.2\pm0.1\%$ and the standard deviation is $\sim$4\%.

The FWHM and line dispersion, $\rm \sigma_{line}$, both used as a proxy for velocity width in the virial mass formula, have their pros and cons. All the samples we used have available FWHM measurements of the H$\beta$ line as its measurement has been more common than $\sigma_{\rm line}$ in catalogs. FWHM is also less sensitive to line wings and blending with narrow lines \citep{Peterson_etal2004}. But, for at least the radio-loud subclass, FWHM is known to correlate with quasar orientation \citep{Wills+Browne1986}. \cite{Bonta_etal2020} show that for H$\beta$ both FWHM and $\sigma_{\rm line}$ provide reasonable proxies, with $\sigma_{\rm line}$ being slightly better. Some other investigations indicate that $\sigma_{\rm line}$ gives better virial mass estimates; for example, \cite{Peterson_etal2004} find a tighter virial correlation when using $\sigma_{\rm line}$, and \cite{Denney_etal2013} find that H$\beta$ and C {\sc iv}-based mass agree better when using $\sigma_{\rm line}$. \cite{Collin_etal2006} compare virial black hole mass based on different line-width measurements with black hole mass from the stellar velocity dispersion of RM AGNs. They find that the $\sigma_{\rm line}$-based mass is better, albeit at low statistical significance. Recently, \cite{Wang_etal2019} presented a similar study using the SDSS-RM quasars and reported that although FWHM suffers from orientation effects more than $\sigma_{\rm line}$, the use of $\sigma_{\rm line}$ does not guarantee a better virial mass estimate. An anti-correlation between FWHM and the virial coefficient is known, and using a constant value for $f$ introduces additional uncertainty to the mass estimates \citep{Restrepo_etal2018}.

If the average viewing angle in the low-$z$ RM sample differs from that in luminous bright high-$z$ quasars, this could introduce an additional bias in the black hole mass estimates. The virial coefficient $f$ very likely correlates with inclination angle \citep{Pancoast+2014, Collin_etal2006}.

A couple of possible competing orientation effects can introduce systematic biases that we may investigate in the future. For example, there is an observational bias due to the anisotropic nature of an accretion disk. The continuum emission varies with orientation \citep[e.g.,][]{Runnoe+2013}: a face-on disk is brighter than a relatively edge-on disk leading to a selection bias towards more face-on sources in luminous high-$z$ samples \citep[][]{DiPompeo+2014}. Although, if the selection is entirely random, we should observe more relatively edge-on sources because of the higher probability of line of sight being edge-on than face-on. These factors must be considered along with the likelihood that AGN opening angles increase with increasing luminosity \citep[e.g.,][]{Lawrence1991,Ma&Wang2013}.  

It is worth noting that the RM mass has an inherent uncertainty of 0.3-0.5 dex due to its calibration against the $M-\sigma_{\star}$ relation \citep{Peterson2010, Vestergaard2011, Shen2013, Ho&Kim2014}. The SE mass estimates have a 0.5-0.6 dex relative uncertainty and 0.7 dex absolute uncertainty \citep[e.g., Table 5,][]{VP2006}.

Our updated SE mass prescription using the $\mathcal{R}_{\rm Fe}$-based correction will not improve the precision of the estimate in individual objects, but will improve the accuracy, particularly for those with high accretion rates, which otherwise would be systematically overestimated.  While the correction will generally be smaller than the overall absolute uncertainty, correcting for systematic effects is important.

\section{Conclusions and Future Outlook}

The recently established R-L relation by \cite{Du_2019} takes into account the bias due to the accretion rate using $\mathcal{R}_{\rm Fe}$. We use quasar samples across a wide range of redshifts from \cite{Shemmer2004}, \cite{Netzer2004}, \cite{Shemmer2010}, \cite{Shen_2011}, \cite{Shen2016} and \cite{Sulentic2017} to characterize the bias in black hole mass and accretion rate when using the canonical and \cite{Du_2019} R-L relationships. The single-epoch black hole mass estimates using the canonical R-L relationship systematically overestimate the mass of the black hole and underestimate the accretion rate. At high redshift, the black hole mass has likely been overestimated by a factor of two on average when using the canonical R-L relationship. The overestimation could be up to an order of magnitude for the most highly accreting quasars, likely the most luminous such objects in the early universe. Our results also indicate that the high-redshift luminous quasars have highly accreting black holes whose optical spectra have characteristically  strong $\mathcal{R}_{\rm Fe}$. The low-redshift analogs of these highly accreting quasars are less massive but also exhibit strong $\mathcal{R}_{\rm Fe}$. The use of the canonical R-L relationship results in an overestimation of black hole mass by a factor of two, on average, for both these highly accreting quasar populations.

The largest galaxy interactions/growth and star formation rates occur at cosmic noon, along with the most luminous quasars that will preferentially have very high accretion rates. \cite{Kelly_Shen2013} show that the fraction of highly accreting AGNs increases with increasing redshift.  The gas fraction in AGN host galaxies increases with redshift, likely contributing to the higher accretion rates \citep{Shirakata_etal2019}. Hence, the mass and accretion rate corrections are relatively common among the most luminous quasars at cosmic noon, and likely the case at even higher redshifts. Low-accretion rate AGNs likely exist at these redshifts but are more commonly below the the flux limits of surveys like SDSS, or only have low S/N spectra currently available \citep[e.g.,][]{Kelly_Shen2013}.

In the absence of rest-frame optical spectra for high-redshift objects, Mg {\sc ii} and C {\sc iv} provide an alternative for black hole mass estimation. Such mass estimates assume that the emission line follows normal ``breathing'', i.e., as continuum luminosity increases, the time lag between continuum and emission-line variation increases, and the emission-line width decreases. Only the H$\beta$ line truly follows this; Mg {\sc ii} shows no breathing, whereas C {\sc iv} shows anti-breathing \citep{Wang_etal2020}. To correctly determine the black hole mass and accretion rate for high-redshift quasars, we need more near-IR spectroscopic surveys to facilitate direct checks using the H$\beta$ line. In the future, the Gemini Near-Infrared Spectrograph-Distant Quasar Survey \citep{Matthews_etal2021} will provide a large, uniformly distributed sample of quasars at high redshift. It will provide measurements of $\rm H\beta$, Fe {\sc ii}, [O {\sc iii}], and other UV lines that fall in the near-IR regions. The final data set will have a high signal-to-noise ratio (SNR$\sim$35) in the observed-frame $\sim$ 0.8 - 2.5 $\mu$m band for a few hundred SDSS quasars at 1.5 $\leq z \leq$ 3.5. The wavelength coverage of the survey may enable the use of the C {\sc iv} line and other UV features to determine an $\mathcal{R}_{\rm Fe}$ equivalent in the near-IR.

The presence of a billion solar mass quasar at $z>6$ is a problem because it is challenging to create and provide constant feeding of the black hole when the universe was less than a billion years old \citep[][]{Turner1991,Haiman+Loeb2001,Shen2013,Inayoshi+2020}. Our results indicate that the black hole masses of high-redshift quasars are typically overestimated, especially for the most highly accreting black holes. Understanding the formation and evolution of massive black holes in the early universe necessitates taking into consideration $\mathcal{R}_{\rm Fe}$-based accretion rate bias in their black hole mass estimates.

\section*{Acknowledgements}
We thank the anonymous referee for the constructive comments that improved the manuscript. JMW acknowledges support from the National Science Foundation of China (NSFC-11833008 and -11991054) and from the National Key R\&D Program of China (2016YFA0400701). PD acknowledges the support from NSFC-11873048, -11991051 and the Strategic Priority Research Program of the Chinese Academy of Sciences (XDB23010400). This work is supported by National Science Foundation grants AST-1815281 (O. S., B. M., C. D.) and AST-1815645.

\section*{Data Availability}

The data underlying this article are available in \citealt{Shen_2011} (DOI: 10.1088/0067-0049/194/2/45), \citealt{Shen2016} (DOI:  
10.3847/0004-637X/817/1/55), \citealt{Shemmer2004} (DOI: 10.1086/423607), \citealt{Netzer2004} (DOI: 10.1086/423608), \citealt{Sulentic2017} (10.1051/0004-6361/201630309), and \citealt{Shemmer2010} (DOI: 10.1088/2041-8205/722/2/L152). See Section 3 for the sample selection criteria applied.


\bibliographystyle{mnras}
\bibliography{Mass_accretion} 




 \appendix

\section{Selection criteria producing the high-z sub-samples}\label{app:app1}
\begin{enumerate}
\item \cite{Shen2016} cataloged the properties of 74 quasars in the redshift range $1.5\leq z\leq3.5$. 60/74 targets are from \citet{Shen&Liu2012} in the redshift range 1.5 to 2.2, and 14 new quasars at z $\sim$ 3.3 were added by \cite{Shen2016}. All targets were selected from the SDSS DR7 quasar catalog to have an S/N $>$10 in the C {\sc iv} through the Mg {\sc ii} spectral regions. They excluded targets with broad absorption features or unusual continuum shapes. Such selection criteria at high redshift result in a sample of very luminous quasars, yet they demonstrate similar diversity in spectral features as seen in low-$z$ quasars \citep[][]{Shen&Liu2012, Shen2016}. Figure 3 of \cite{Shen2016} presents the median spectrum of these 74 quasars. The median spectrum shows broader H$\beta$ and H$\alpha$ lines characteristic of massive black holes as compared to the low-z quasars. These quasars span nearly the entire range of $\mathcal{R}_{\rm Fe}$, and follow the EV1 trends seen in the low-z quasars \cite[][Figures 7 \& 8]{Shen2016}.

\item The SN2004 sample includes 29 quasars in the redshift range $2\leq z \leq3.5$. They selected luminous quasars ($L\geq 10^{46} \rm erg~ s^{-1}$) with H magnitudes $\leq 17$ allowing them to obtain high S/N IR spectra. These quasars had archival UV spectra including the N {\sc v} and C {\sc iv} emission lines without severe absorption. Their selection criteria also required the H$\beta$ line to be unaffected by atmospheric absorption in the IR bands. \cite{Shemmer2004} show the spectrum of these quasars in their Figures 1, 2 \& 3. These high-z quasars also follow the EV1 trend, in fact, they occupy the same region on the $\mathcal{R}_{\rm Fe}$ vs $L_{\rm Bol}/L_{\rm Edd}$ plot as the narrow-line Seyfert 1 galaxies from \citealt{BG1992} \citep[][Figure 8]{Netzer2004}.

\item \cite{Sulentic2017} selected 28 quasars from the magnitude-limited ($m_{\rm B}\approx 17.5)$ Hamburg ESO survey \citep[][]{Wisotzki+2000} with $z>1.4$ to allow observation of the C {\sc iv} spectral region. They are high-luminosity (log$L_{5100 \angstrom} > 46~ \rm erg~ s^{-1}$) quasars with the H$\beta$ region properties reported by \citealt{Sulentic+2004,Sulentic+2006, Marziani+2009}. Two quasars in \cite{Sulentic2017} are known to be gravitationally lensed, another is a mini-broad absorption line quasar, and two are weak-line quasars (WLQs).
 \end{enumerate}

\section{Weak-line quasars}\label{app:app3}
With the assumption that virial mass method holds true for WLQs, we show the effect of the Fe-correction on the determination of their black hole mass. Our analysis includes four WLQs, two from \cite{Sulentic2017} and two from \cite{Shemmer2010}. 

We discussed the sample selection criterion of \cite{Sulentic2017} in Appendix \ref{app:app1}; their WLQs have redshifts 1.52 and 1.58. WLQs in the \cite{Shemmer2010} sample are at a redshift of 3.49 and 3.55 and were selected from \cite{Collinge+2005} for near-IR spectroscopy as their H$\beta$ regions fall in the middle of the K band. These latter WLQs show weaker H$\beta$ lines \citep[][Figure 2]{Shemmer2010} in comparison to high-z quasars with similar luminosities, but have reliable measurements of the parameters used for black hole mass estimates \citep[][]{Shemmer2010}. \cite{Shemmer2010} suggest the weakness of low and high-ionization emission lines is due to a gas deficiency in the BLR indicated by a low BLR covering factor rather than an effect of extreme accretion rate. The H$\beta$ lines in WLQs from \citealt{Sulentic2017} are of normal strength.

\begin{figure*}
   \begin{subfigure}{0.5\textwidth}
     \centering
   \includegraphics[trim=0.cm 0.cm 0.5cm 2.cm,clip,width=0.95\columnwidth]{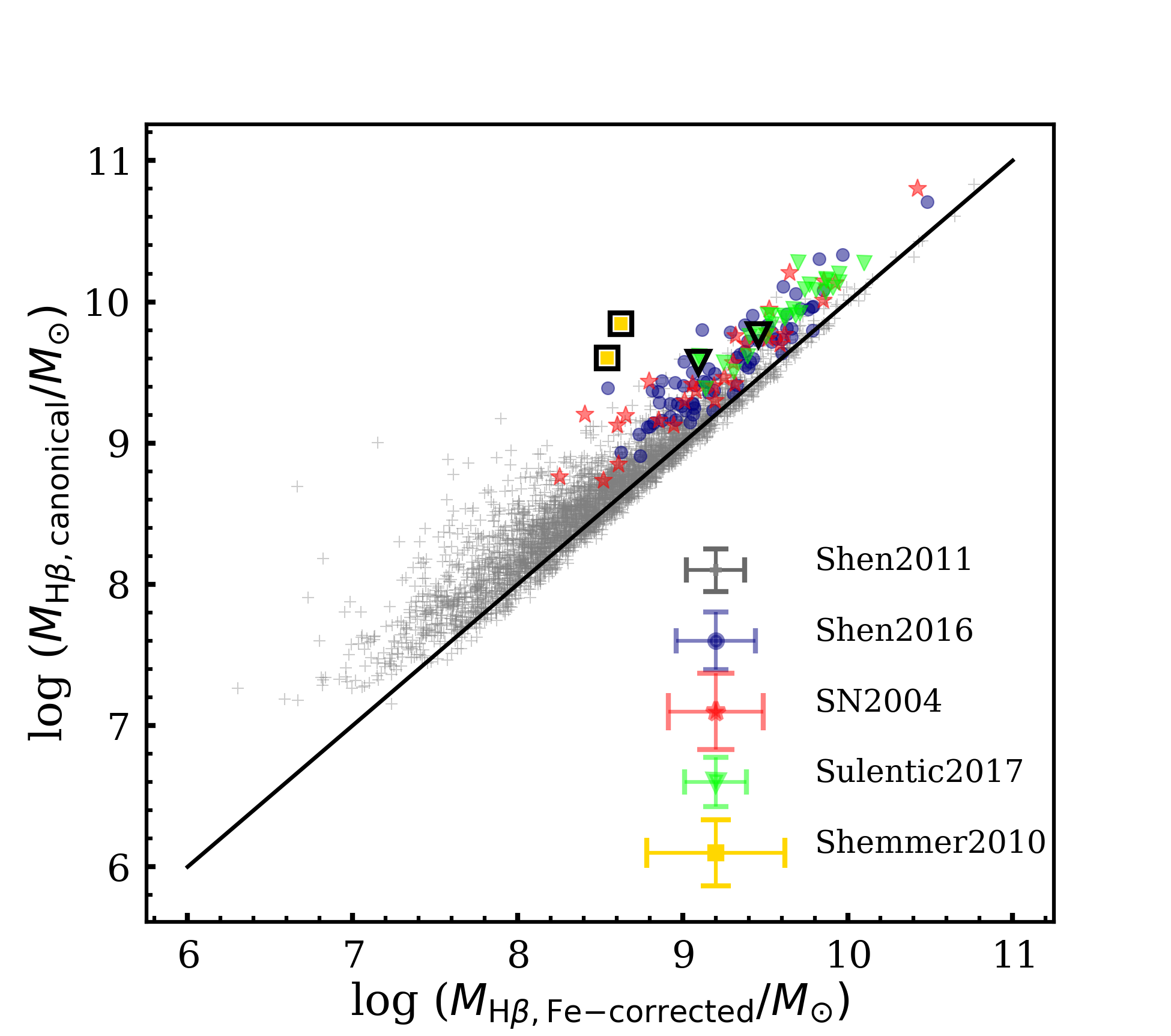}
   \end{subfigure}%
   \centering
   \begin{subfigure}{0.5\textwidth}
     \centering
     \includegraphics[trim=0.9cm .5cm 1.5cm 2.cm,clip,width=1.05\columnwidth]{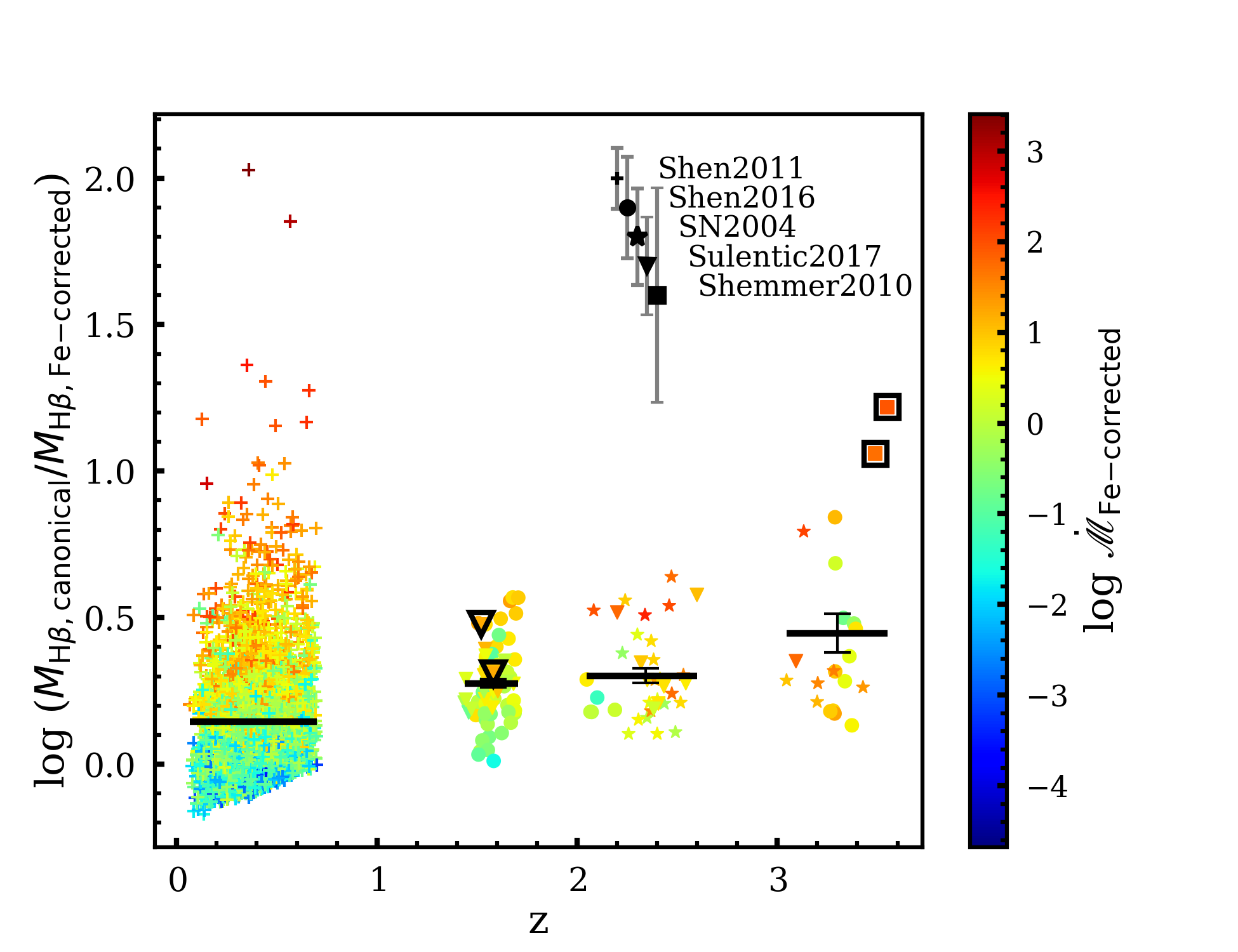}
   \end{subfigure}%
   \caption{Left: Black hole mass calculated using the canonical (y-axis) and Fe-corrected (x-axis) R-L relations. Right: Mass ratio between canonical and Fe-corrected black hole mass vs redshift colour-coded by accretion rate. In both plots, WLQs are outlined in black. }
   \label{fig:AppB}
\end{figure*}

After utilizing the \cite{Du_2019} R-L relationship, Figure \ref{fig:AppB} (left) shows that the two WLQs from \cite{Shemmer2010} sample have extreme accretion rates, and their mass overestimated by a factor of $\sim$14. The two WLQs from \cite{Sulentic2017} show a mass overestimation of a factor of $\sim$2.5. Although Fe-correction is crucial for all four WLQs as they have strong Fe {\sc ii }, the effect of Fe-correction is most extreme in the WLQs with weaker H$\beta$ lines ($\mathcal{R}_{\rm Fe} \propto 1/{\rm EW~ H\beta}$) from \cite{Shemmer2010} in comparison to the WLQs from \cite{Sulentic2017}. However, WLQs with weak H$\beta$ lines in \cite{Shemmer2010} have larger measurement uncertainties associated with their FWHM and EW of H$\beta$ measurements and hence larger measurement errors in masses. The mass ratio vs redshift plot (Figure \ref{fig:AppB} right) does not have a significant change in mean mass ratio in the redshift bin by the inclusion of just two WLQs. The mean mass differences are 0.14$\pm$0.00 dex for $z<1$, 0.28$\pm$0.01 dex for $1<z<2$, 0.30$\pm$0.02 dex for $2<z<3$ and 0.45$\pm$0.07 dex for $z>3$.

\section{SDSS spectra of highly accreting quasars in Shen et al. 2011}\label{app:app2}

Table \ref{tab:append} lists the name, EW of H$\beta$, EW of Fe {\sc ii}, $\mathcal{R}_{\rm Fe}$, and log of mass ratio i.e., log($M_{\rm H\beta, canonical}/ M_{\rm H\beta, Fe-corrected}$) of \cite{Shen_2011} quasars that appear as outliers (log of mass ratio $>1$) in Figures \ref{fig:Figure6a}, \ref{fig:Figure6} \& \ref{fig:Figure7}. 
Figure \ref{fig:Figure B1} shows the SDSS spectra of these quasars arranged in descending order of mass ratio. SDSS J152350.42+391405.2 shows broad absorption line features.

\begin{table*}
 \caption{Outliers in Shen et al. (2011) sample}
 \label{tab:append}
 \begin{tabular}{llcclcl}
 \hline
 SDSSJ & z & EW(H$\beta$) & EW(Fe{\sc ii}) & $\mathcal{R}_{\rm Fe}$ & log(Mass ratio) \\
    &  &  (\angstrom)   &   (\angstrom)   &   &  (dex) \\
 \hline
J140325.82+443014.1 & 0.4122 & 28.6 & 89.3  & 3.12 & 1.02 \\    
J094704.51+472142.8 & 0.5392 & 26.9 & 81.1  & 3.01 & 1.03 \\    
J131609.78-015403.9 & 0.4057 & 14.0 & 43.3  & 3.09 & 1.03 \\    
J093531.60+354101.0 & 0.4936 & 23.0 & 77.2  & 3.36 & 1.16 \\    
J104431.76+070841.0 & 0.6477 & 19.8 & 66.7  & 3.37 & 1.17 \\    
J224028.85-010649.8 & 0.1268 & 18.9 & 70.6  & 3.74 & 1.18 \\    
J152350.42+391405.2 & 0.6609 & 48.9 & 175.6 & 3.59 & 1.28 \\    
J130601.87+580319.9 & 0.4437 & 10.2 & 39.0  & 3.82 & 1.31 \\    
J092153.63+033652.6 & 0.3509 & 25.7 & 106.3 & 4.14 & 1.36 \\    
J083525.98+435211.3 & 0.5676 & 18.5 & 99.0  & 5.35 & 1.85 \\    
J131549.46+062047.8 & 0.3600 & 21.7 & 129.1 & 5.95 & 2.03  \\   
 \hline
 \end{tabular}
\end{table*}

\begin{figure*}
\begin{center}
\includegraphics[trim=0.cm 0.0cm 0.cm 0.5cm,clip,width=2.\columnwidth]{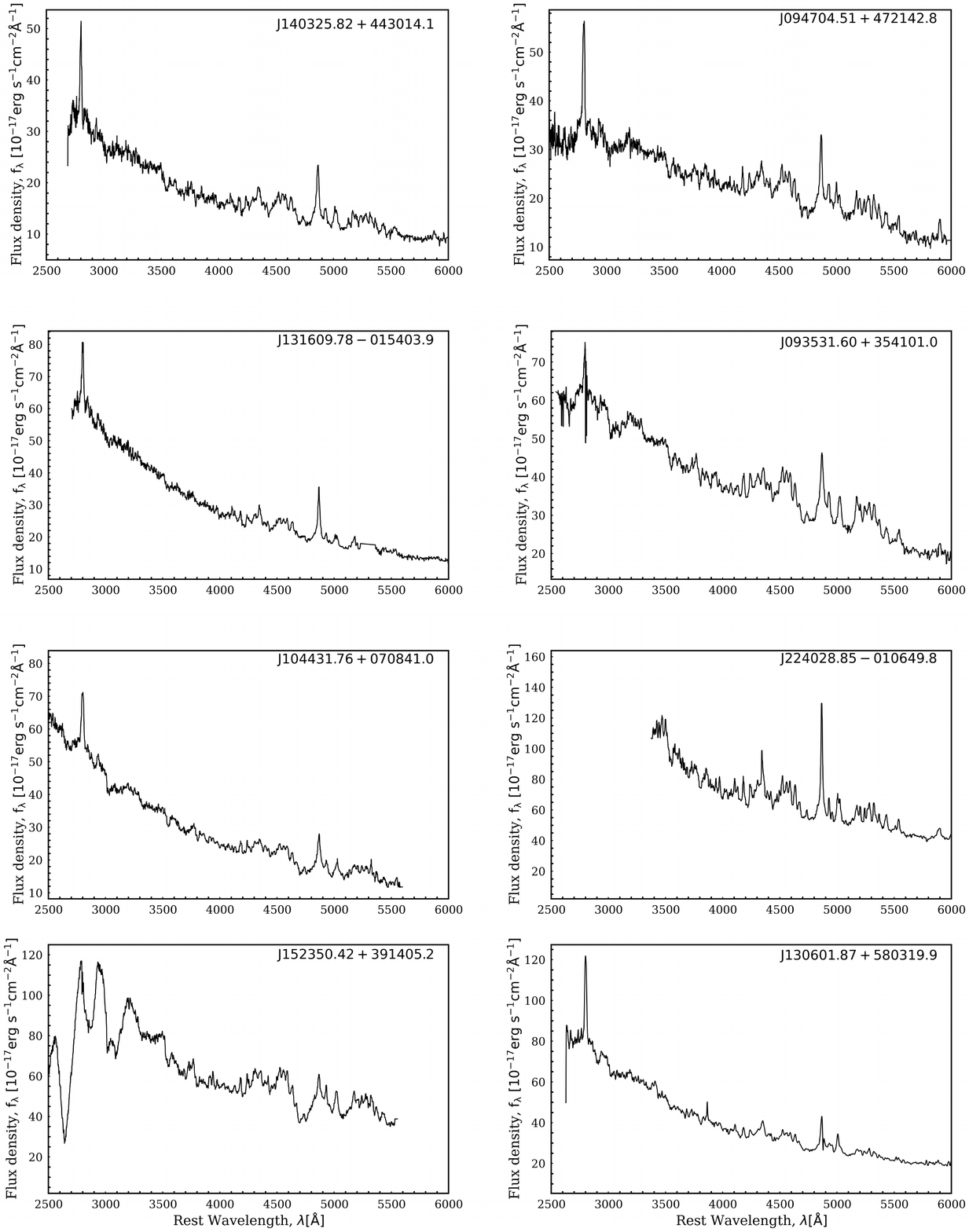}
\caption{SDSS spectrum of outliers in Shen et al. (2011) sample. They have $\mathcal{R}_{\rm Fe}>3$ and log ($M_{\rm H\beta, canonical}/ M_{\rm H\beta, Fe~ corrected}) \geq 1$. }
\label{fig:Figure B1}
\end{center}

\end{figure*}

\begin{figure*}
\begin{center}
\includegraphics[trim=0.cm 14.0cm 0.cm 0.0cm,clip,width=2.\columnwidth]{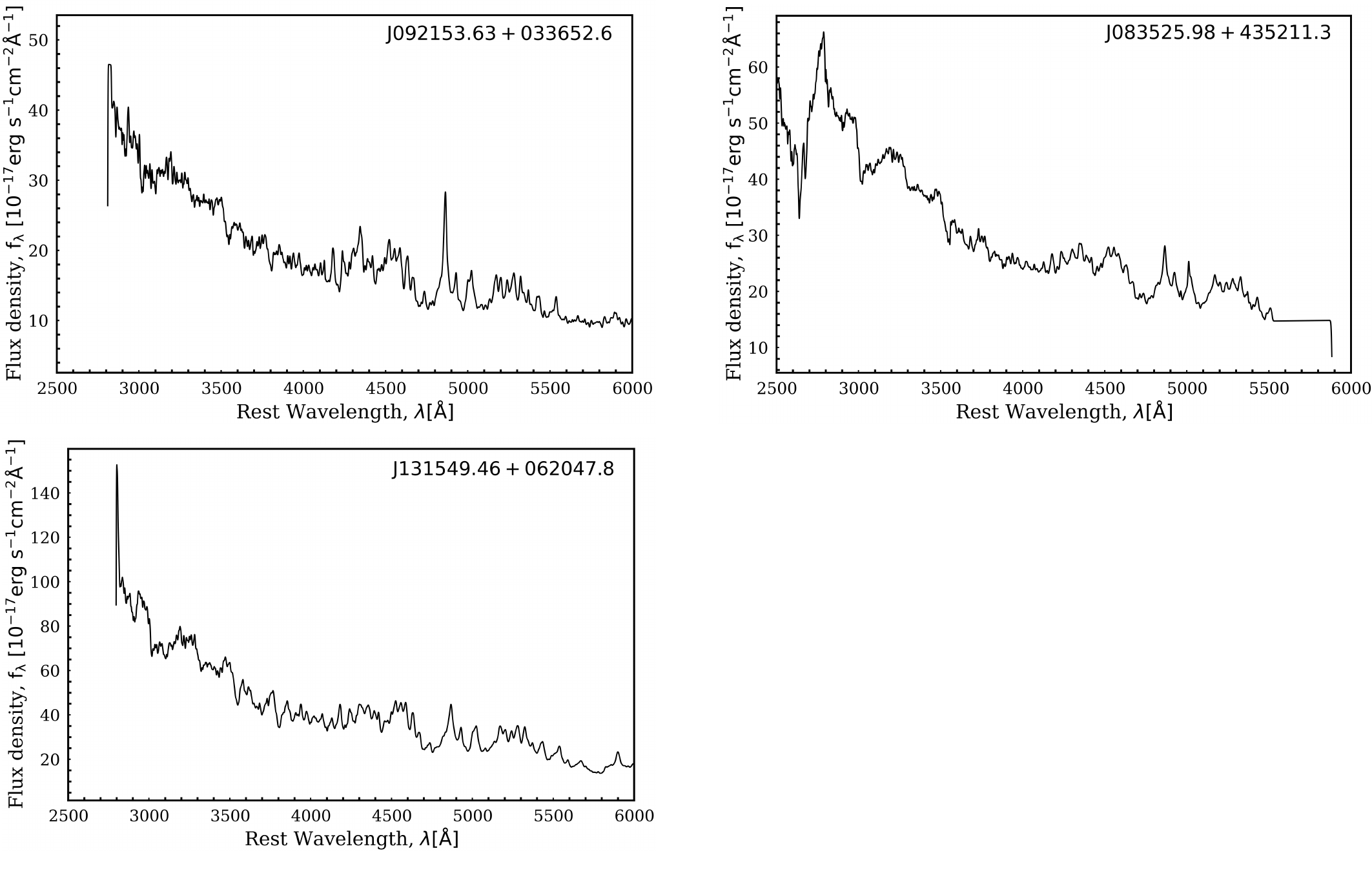}
\contcaption{}
\label{fig:FigureB1 cont}
\end{center}
\end{figure*}

\bsp	
\label{lastpage}
\end{document}